\documentclass[graybox]{svmult}

\usepackage{mathptmx}       
\usepackage{helvet}         
\usepackage{courier}        
\usepackage{type1cm}        

\usepackage{makeidx}         
\usepackage{graphicx}        
\usepackage{multicol}        
\usepackage[bottom]{footmisc}

\usepackage{enumitem}
\usepackage{setspace}
\usepackage{amsmath}
\usepackage{amssymb}
\usepackage{subfigure}
\usepackage{amsfonts}
\usepackage{dsfont}
\usepackage{ifpdf}
\usepackage{eurosym}
\usepackage{xcolor} 
\usepackage{placeins}
\usepackage{psfrag}
\usepackage{epsfig}
\usepackage{float}

\makeindex             

\newcommand{\eq}[1]{Eq.~(\ref{#1})}    
\newcommand{\fg}[1]{Fig.~\ref{#1}}     

\newcommand \be  {\begin{equation}}
\newcommand \bea {\begin{eqnarray} \nonumber }
\newcommand \ee  {\end{equation}}
\newcommand \eea {\end{eqnarray}}

\graphicspath{{./figs/}}  

\begin{document}

\title*{Study of statistical correlations in intraday and daily financial return time series}
\titlerunning{Statistical correlations in financial return time series}
\author{Gayatri Tilak, Tam\'{a}s Sz\'{e}ll, R\'{e}my Chicheportiche and Anirban Chakraborti}
\authorrunning{G. Tilak, T. Sz\'{e}ll, R. Chicheportiche and A. Chakraborti}
\institute{Gayatri Tilak \at Universit\'{e} Paris Dauphine, Place du Mar\'{e}chal de Lattre de Tassigny 75016 Paris, France \\
Laboratoire MAS, \'{E}cole Centrale Paris, 92290 Ch\^{a}tenay-Malabry, France \\ \email{gayatri.tilak@ecp.fr}
\and Tam\'{a}s Sz\'{e}ll \at Budapest University of Technology and Economics \\ Muegyetem rkp. 3-9. H-1111 Budapest, Hungary \\ \email{tomi@digiflex.hu}
\and R\'{e}my Chicheportiche \at 
Laboratoire MAS, \'{E}cole Centrale Paris, 92290 Ch\^{a}tenay-Malabry, France \\ \email{remy.chicheportiche@ecp.fr}
\and Anirban Chakraborti \at 
Laboratoire MAS, \'{E}cole Centrale Paris, 92290 Ch\^{a}tenay-Malabry, France \\ \email{anirban.chakraborti@ecp.fr}
}
%
%
\maketitle

\abstract{The aim of this article is to briefly review and make new studies of correlations and co-movements of stocks, so as to understand the ``seasonalities'' and market evolution. Using the intraday data of the CAC40, we begin by reasserting the findings of Allez and Bouchaud \cite{Allez}: the average correlation between stocks increases throughout the day. We then use multidimensional scaling (MDS) in generating maps and visualizing the dynamic evolution of the stock market during the day. We do not find any marked difference in the structure of the market during a day. Another aim is to use daily data for MDS studies, and visualize  or detect specific sectors in a market and periods of crisis. We suggest that this type of visualization may be used in identifying potential pairs of stocks for ``pairs trade''.}

\section{Introduction}
\label{intro}
Many complex features, including multi-fractal behaviour, of financial markets have been studied for a long time, and constitute today a collection of empirical ``laws'', the so-called ``stylized facts'' \cite{munitoke}. The questions: ``How efficient is the market? To what extent?'' have been long debated on by economists, econometricians and practitioners of finance \cite{mp}. It is now accepted that the market is weakly efficient (at least to some extent and in certain time scales), and that several quantities like the price returns, volatility, traded volume, etc. do exhibit ``seasonal patterns''\footnote{``The existence of seasonal asset returns may be an indicator of market inefficiencies\ldots The presence of seasonal returns, however, does not necessitate market inefficiency''\cite{JSTOR}}; why these ``market anomalies'' appear is, of course, not well-understood. One reason for their appearance could be that the markets operate in synchronization with human activities and so the financial time series of returns of many assets reveal the related statistical ``seasonalities''. Identifying such anomalies in order to make statistical arbitrage is a usual practice. Another related practice is estimating market co-movements, which is certainly relevant in several areas of finance, including investment diversification \cite{Markowitz} and risk management \cite{Embrechts}. 

In this paper, we first present some notations, definitions and methods. We then review existing results on intraday patterns concerning both individual and collective stock dynamics. We compare the cross-sectional ``dispersion'' of returns and its typical evolution during the day, with the intraday pattern of the  leading modes  of the cross-correlation matrix between stock returns, following the studies of Allez and Bouchaud \cite{Allez}. Then, we make additional plots of the pair-wise 
cross-correlation matrix elements and study their typical evolution during the day.
Finally, we  use multidimensional scaling (MDS) in generating maps and visualizing the dynamic evolution of the stock market during the day. When the MDS studies are repeated  with daily data, we find that it is easier to visualize  or detect specific sectors and market events. We suggest that this type of plots may be used in identifying potential pairs of stocks for ``pairs trade''. 

\section{Some data specifications, notations, and definitions}

In order to measure co-movements in the time series of stock prices, the popular Pearson correlation coefficient is commonly used. However, it is now known that several factors viz., the statistical uncertainty associated with the finite-size time series, heterogeneity of stocks, heterogeneity of the average inter-transaction times, and asynchronicity of the transactions may affect the reliability of this estimator. The investigation of high-frequency ``tick-by-tick'' data does enable one to monitor market co-movements and price formation in real time. However, high-frequency data have the drawback of aggravating the above mentioned factors even further, raising the need to adequately evaluate their impact through proper correlation measures, such as the Hayashi-Yoshida estimator \cite{HY}. In this section, we introduce such concepts, along with notations and definitions, and also specify the details of the datasets used.

We have considered three data sets.
\begin{itemize}
\item{Daily returns:} we have used the freely downloadable daily closure prices from Yahoo for $N=54$ companies in the New York Stock Exchange, over a period spanning from January 1, 2008 to May 31, 2011.

\item{Intraday tick-by-tick:} $N=40$ companies of the CAC40 stock exchange for March 2011, between 10:00-16:00 CET. We have purposefully avoided the opening and closing hours of the market, so as to avoid certain anomalies.

\item{Intraday sampled retuns:} Same universe as the tick-by-tick but sampled in bins of $5$ minutes or $30$ minutes.  Thus, the total number of $5$ minute bins is $72$ per day and total number of $30$ minute bins is $12$ per day. The total number of trading days in one month is around $T=21$.

\end{itemize}

\subsection{Cross-sectional ``dispersion'' of the binned data}
In this section we introduce the notations and definitions used by the authors of  Ref.~\cite{Allez} for their study of sampled intraday data; we will use the same notations when reproducing their results for our own dataset.

Stocks are labelled by $i=1,\dots,N$, days by $t=1, \dots, T$ and bins by $k=1, \dots, K$. The return of stock $i$ in bin $k$ of day $t$ will be denoted as $r_i(k;t)$. The temporal distribution of stock $i$ in bin $k$ is characterised by its moments: mean $\mu_i(k)$ and standard deviation (volatility) $\sigma_i(k)$,
which are defined as:
\begin{subequations}\label{single}
\begin{align}
	\mu_i(k)&= \langle r_i(k;t) \rangle\\
	\sigma_i^2(k)&=\langle r_i(k;t)^2 \rangle - \mu_i^2(k),
\label{kurtosis}
\end{align}
\end{subequations}
where averages over days for a given stock and a given bin are expressed with angled brackets: $\langle \dots \rangle$.

The cross-sectional ``dispersion'' of the returns of the $N$ stocks for a given bin $k$ in a given day $t$ is as well characterised by its moments:
\begin{subequations}
\begin{align}
	\mu_d(k;t)&= [r_i(k;t)]\\
	\sigma_d^2(k;t)&=[ r_i(k;t)^2 ] - \mu_d^2(k;t),
\end{align}
\label{collective}
\end{subequations}
where
the averages over the ``ensemble'' of stocks for a given bin in a given day are expressed with square brackets: $[ \dots ]$.
We note that $\mu_d(k;t)$ may be interpreted as the ``return of an index'', equiweighted on all stocks. We will be more interested in the average of $ \sigma_d^2(k;t) $ over all days, as a way to characterise the typical intraday evolution of the ``dispersion'' between stock returns. Detailed studies of this dispersion and other such measures, concerning both stock prices and returns, will be presented elsewhere \cite{Esteban}.

Although the dispersion, described above, indicates the ``co-movements'' of stocks, a more common and direct characterisation is through the standard ``correlation'' of returns. In order to measure the correlation matrix of the returns, each return is \textit{normalised} by the dispersion of the corresponding bin, to reduce the intraday seasonality and also take into account the fluctuation of the volatility in the considered time period $ T $. Therefore, following the same prescription as in Ref.~\cite{Allez}, we define: $\widehat r_i(k;t)=r_i(k;t)/\sigma_d(k;t)$ and study the correlation matrix defined for a 
given bin $k$:
\be
\label{correlation}
\rho_{ij}(k) := \frac{\langle \widehat r_i(k;t) \widehat r_j(k;t) \rangle-\langle \widehat r_i(k;t) \rangle\langle \widehat r_j(k;t) \rangle}{\widehat \sigma_i(k)\widehat \sigma_j(k)} .
\ee
The largest eigenvalue of the $N\times N$ correlation matrix $\mathbf{C}(k)$ composed of the elements $\rho_{ij}(k)$, is denoted by $\lambda_1(k)$ and is equal to the risk of the corresponding eigenmode, the ``market mode'' with all entries positive and close to $1/\sqrt{N}$. In fact, $\lambda_1(k)/N$ can be seen as a measure of the average correlation between stocks. We will be interested in the intraday evolution or the bin-dependence of the largest eigenvalue\footnote{A similar study about the intraday evolution of the first eigenvector is of great interest and has been performed as well in \cite{Allez}.}.

\subsection{Correlation matrix with tick-by-tick data}
Computing correlations using these intraday data, raises lots of issues concerning usual estimators, as already indicated above.
Let us assume that we observe $T$ time series of prices or log-prices $p_i, ({i=1,\ldots,T})$, observed at times $t_m ({m=0,\ldots,M})$. The usual estimator of the covariance of prices $i$ and $j$ is the \emph{realized covariance estimator}, which is computed as:
\begin{equation}
	\hat\Sigma^{RV}_{ij}(t)=\sum_{m=1}^M (p_{i}(t_m)-p_{i}(t_{m-1}))(p_{j}(t_m)-p_{j}(t_{m-1})). \nonumber
\end{equation}

The problem is that high-frequency tick-by-tick data record changes of prices when they happen, i.e. at times not predefined and not equidistant. Multivariate tick-by-tick data are thus asynchronous, contrary to daily close prices for example, which are by construction synchronous for all the assets on a given exchange. Using standard estimators without caution, could be one cause for the ``Epps effect'', first observed in \cite{Epps1979}, which stated that ``correlations among price changes in common stocks of companies in one industry are found to decrease with the length of the interval for which the price changes are measured.'' Hence, here we use the Hayashi-Yoshida estimator \cite{HY} also, which takes (part of) the Epps effect into account. There are many other estimators that may be used in general, and a comparison of such estimators has been performed in Ref. \cite{Giulia}.

\subsubsection*{  Hayashi-Yoshida (HY) estimator} 

In \cite{HY}, the authors introduced a new estimator for the linear correlation coefficient between two asynchronous diffusive processes. Given two It\^{o} processes $X,Y$ such that
\begin{align}
&{\rm d}X_t=\mu^X_t{\rm d}t+\sigma^X_t{\rm d}W^X_t\\
&{\rm d}Y_t=\mu^Y_t{\rm d}t+\sigma^Y_t{\rm d}W^Y_t\\
&{\rm d}\left\langle W^X,W^Y\right\rangle_t=\rho_t{\rm d}t,
\end{align}
and observation times $0=t_0\leq t_1\leq\ldots\leq t_{n-1}\leq t_n=T$ for $X$, and $0=s_0\leq s_1\leq\ldots\leq s_{m-1}\leq s_m=T$ for $Y$, which must be independent for $X$ and $Y$, they showed that the following quantity:\\
\begin{align}
&\sum_{i,j}r^X_ir^Y_j\mathds{1}_{\left\{O_{ij}\neq \emptyset \right\}},\\
&O_{ij}=]t_{i-1},t_{i}]\cap]s_{j-1},s_{j}] \nonumber \\
&r^X_i=X_{t_{i}}-X_{t_{i-1}} \nonumber\\
&r^Y_j=Y_{s_{j}}-Y_{s_{j-1}}, \nonumber
\end{align}
is an unbiased and consistent estimator of $\int_{0}^T\sigma^X_t\sigma^Y_t\rho_t{\rm d}t$, as the largest mesh size goes to zero.
In practice, it amounts to \textit{summing every product of increments as soon as they share any overlap of time}. In the case of constant volatilities and correlation, it provides a consistent estimator for the correlation
  
\be
\rho_{ij}^t=\frac{\sum_{i,j}r^X_ir^Y_j\mathds{1}_{\left\{O_{ij}\neq \emptyset \right\}}}{\sqrt{\sum_{i}(r^X_i)^2\sum_{j}(r^Y_j)^2}}.
\label{HY}
\ee

\subsection{Pearson correlation coefficient and correlation matrix with daily returns}

In order to study the \textit{equal time} cross-correlations between $ N $ stocks, we first denote the closure price of stock $i$ in day $\tau$ by $P_{i}(\tau)$, and determine the logarithmic return of stock $i$ as 
$r_{i}(\tau)=\ln P_{i}(\tau)-\ln P_{i}(\tau-1)$. For the sequence of $ T $ consecutive trading days, encompassing a given window $t$ with width $ T $, these returns form the \textbf{return vector} $\boldsymbol r_{i}^t$. In order to characterize the synchronous time evolution of assets, we use the equal time Pearson correlation coefficients between assets $i$ and $j$ defined as

\begin{equation}
\rho _{ij}^t=\frac{\langle \boldsymbol r_{i}^t \boldsymbol r_{j}^t \rangle -\langle \boldsymbol r_{i}^t \rangle \langle \boldsymbol r_{j}^t \rangle }{\sqrt{[\langle {\boldsymbol r_{i}^t}^{2} \rangle -\langle \boldsymbol r_{i}^t\rangle ^{2}][\langle {\boldsymbol r_{j}^t}^{2} \rangle -\langle \boldsymbol r_{j}^t \rangle ^{2}]}},
\label{eq:coeff}
\end{equation}

\noindent where $\left\langle ...\right\rangle $ indicates a time average over the $ T $ consecutive trading days included in the return vectors. These correlation coefficients fulfill the usual condition of $-1\leq \rho _{ij}\leq 1$ and form an $N\times N$ correlation matrix $\mathbf{C}^t$, which serves as the basis of further analyses \cite{onnela,ac}.

For analysis, the data is divided time-wise into $M$ \emph{windows} ($t=1,\, 2,...,\, M$) of width $T$, corresponding to the number of daily returns included in the window. The consecutive windows may be overlapping/non-overlapping with each other, the extent of which is dictated by the window step length parameter $\delta t$, describing the displacement of the window, measured also in trading days. The sizes of window width $T$, and window step width $\delta t$, are to be chosen cleverly: for example, $T$ must be long enough to grasp any signal with a certain statistical power, but not cover too long a period over which the signal could have varied. 

\subsection{Distance matrix}
To obtain ``distances'', a non-linear transformation
\be 
d_{ij}=\sqrt{2(1-\rho _{ij})},
\label{distance}
\ee
is used, with the property $2\geq d_{ij}\geq 0$, forming an $N\times N$ distance matrix $\mathbf{D}^t$, such that all distances are ``ultrametric''. The concept of ultrametricity is discussed in detail by Mantegna \cite{Man1}. Out of the several possible ultrametric spaces, the subdominant ultrametric is opted for due to its simplicity and remarkable properties. The choice of the non-linear function is again arbitrary, as long as all the conditions of ultrametricity are met.

\subsection{Multidimensional scaling (MDS)}
Multidimensional scaling is a set of data analysis techniques that display the structure of ``distance''-like data as a ``geometrical picture'', where each object is represented by a point in a multidimensional space. The points are arranged in this space, such that the distances between pairs of points have the strongest possible relation to the ``similarities'' among the pairs of objects --- two similar objects are represented by two points that are close together, and two dissimilar objects are represented by two points that are far apart. The space is usually a two- or three-dimensional Euclidean space, but may be non-Euclidean and may have more dimensions.

MDS is a generic term that includes many different types---classified according to whether the similarities data are ``qualitative'' (called non-metric MDS) or ``quantitative'' (metric MDS). The number of similarity matrices and the nature of the MDS model can also classify MDS types. This classification yields classical MDS (one matrix, unweighted model), replicated MDS (several matrices, unweighted model), and weighted MDS (several matrices, weighted model).
For a general introduction and overview, please see Ref. \cite{MDS}.

The collection of objects to be analyzed in our case, is $ N $ stocks, on which a distance function is defined using \eq{distance}. These distances are the entries of the similarity matrix

\be
    \mathbf{D}^t := \begin{pmatrix} d_{11} & d_{12} & \cdots & d_{1N} \\ d_{21} & d_{22} & \cdots & d_{2N} \\ \vdots & \vdots & & \vdots \\ d_{N1} & d_{N2} & \cdots & d_{NN} \end{pmatrix}.     
\ee

Given $ \mathbf{D}^t $, the aim of MDS is to find $ N $ vectors $x_1,\ldots,x_N \in \mathbb{R}^D $, such that

\be
    \|x_i - x_j\| \approx d_{ij} \quad \forall i,j\in N,
\ee
where $\|\cdot\|$ is a vector norm. In classical MDS, this norm is typically the Euclidean distance metric.

In other words, MDS tries to find a mathematical embedding of the $ N $  objects into $ \mathbb{R}^D $ such that distances are preserved. If the dimension $D$ is chosen to be 2 or 3, we are able to plot the vectors $x_i$ to obtain a visualization of the similarities between the $N$ objects. It may be noted that the vectors $x_i$ are \textit{not unique}-- with the Euclidean metric, they may be arbitrarily \textit{translated} and \textit{rotated}, since these transformations do not change the pairwise distances $\|x_i - x_j\|$.

There are various approaches to determining the vectors $x_i$. Generally, MDS is formulated as an \textit{optimization} problem, where $(x_1,\ldots,x_N)$ is found as a minimization of some cost function, such as

\be
    \min_{x_1,\ldots,x_N} \sum_{i<j} ( \|x_i - x_j\| - d_{ij} )^2. \, 
\ee

A solution may then be found by numerical optimization techniques. In our case, we used simulated annealing as the optimization procedure.

\section{Results}

\subsection{U-effect in volatility}

\begin{figure} 
		\center
		\includegraphics[scale=0.75]{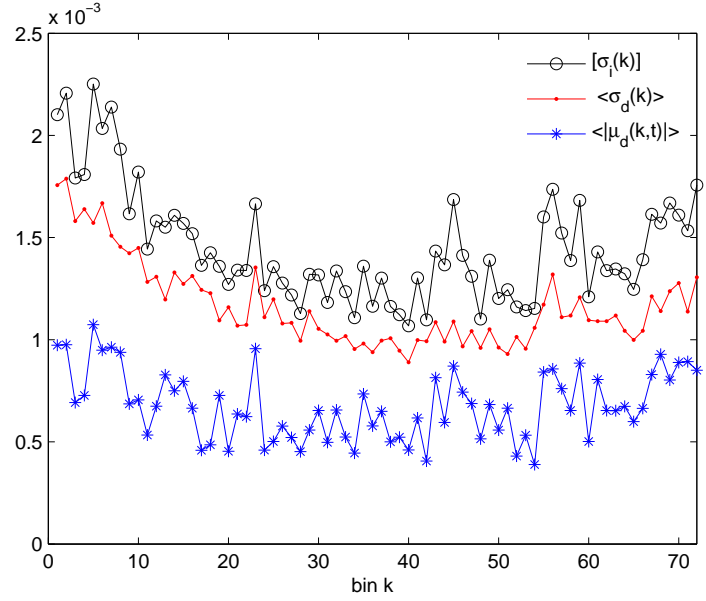}
		\caption{Plots of the average volatility of stocks $\sigma(k)$, the average cross sectional dispersion $\sigma_d(k)$ and 
		the average absolute value of the index return $\langle |\mu_d(k,t)|\rangle$ as a function of the 5-minute bins denoted by $k$, from 10h00-16h00 CET, for the period March, 2011. Courtesy: E. Guevara H. \textit{et al} \cite{Esteban}.
		} 
        \label{figure1}
\end{figure}

In financial studies, among the periodicities or ``seasonalities'' is the ``U-effect'' \cite{admati88,andersen97}, which describes the intraday pattern of average volatility $\sigma(k)=[\sigma_i(k)]$ of individual stocks: the average volatility is high during the market opening hours, then decreases so as to reach a minimum around lunch time, and increases again steadily until the market closes. We show a similar result in Fig.~\ref{figure1}, computed with the CAC40 intraday data for the period March, 2011. The average of $|\mu_d(k;t)|$ is a proxy for the ``index volatility'', and is displayed in Fig.~\ref{figure1} : it also shows a
U-shaped pattern similar to that of $\sigma(k)$.

\subsection{The eigenvalues of the correlation matrix and average correlations}

\begin{figure} 
		\center
		\includegraphics[scale=0.75]{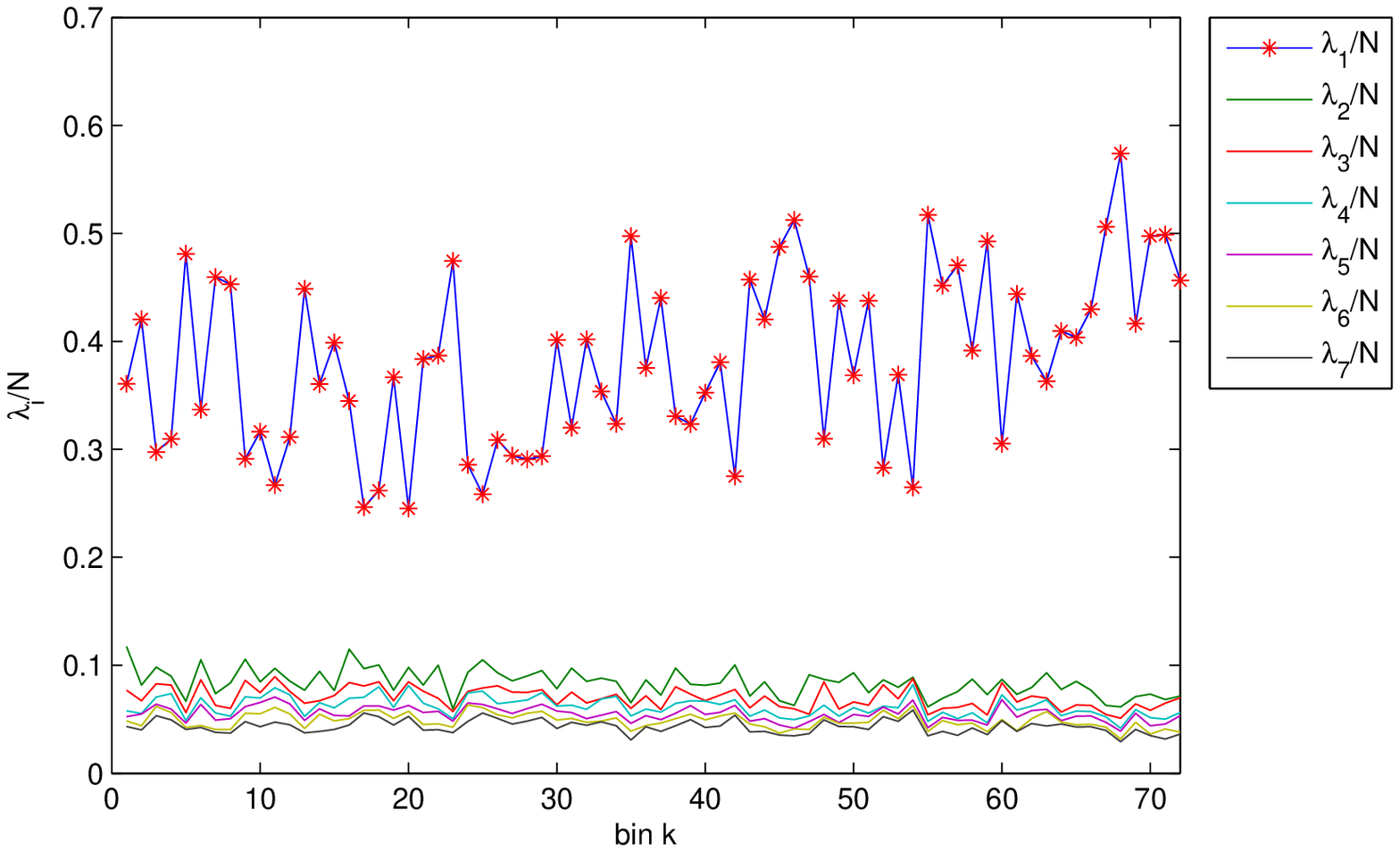}
		\includegraphics[scale=0.35]{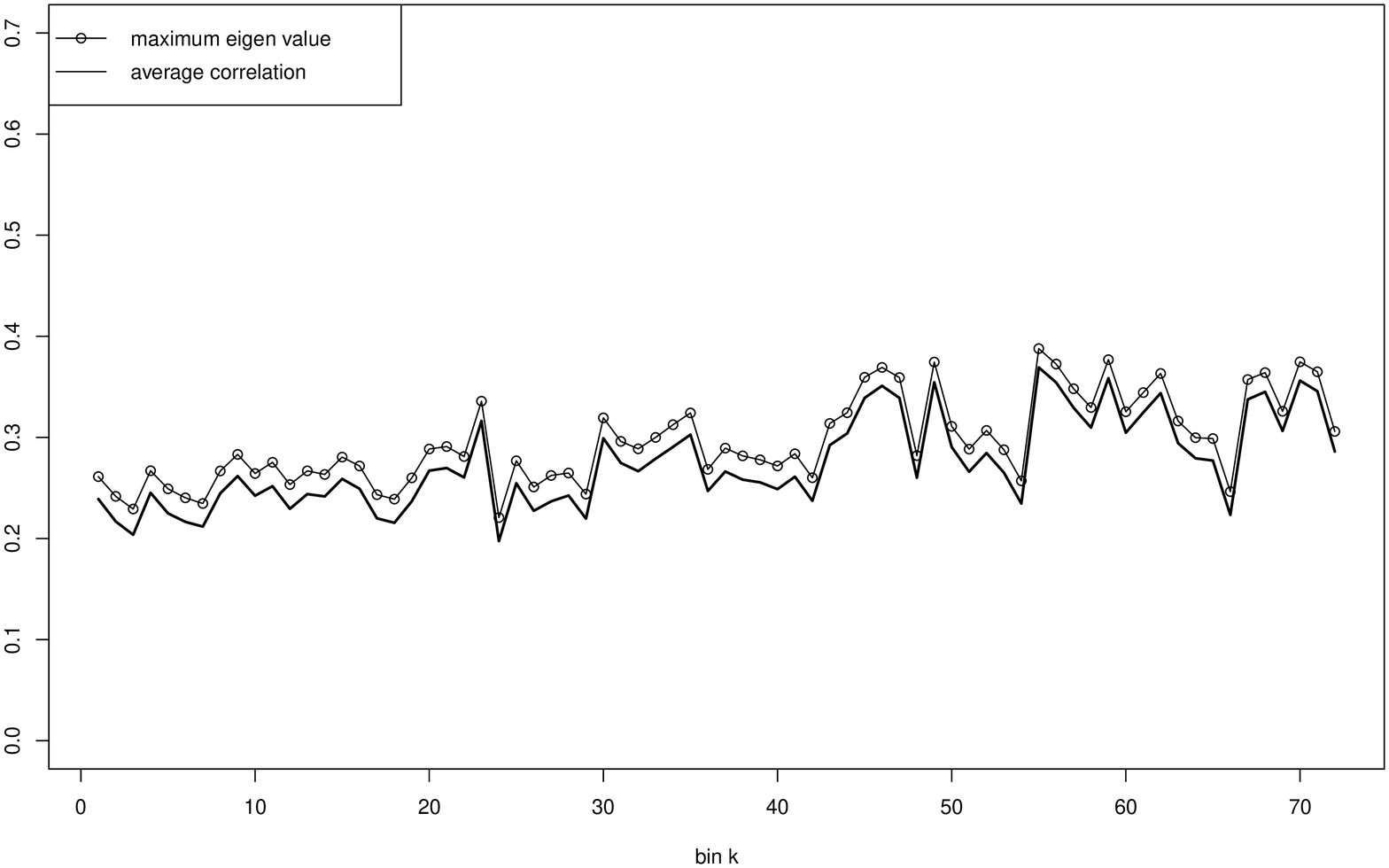}
		\caption{Top: Top eigenvalues of the correlation matrix, $\lambda_i(k)/N$, $i=1, \dots, 7$, as a function of the 5-minute bins denoted by $k$, from 10h00-16h00 CET, in March, 2011. [5-min sampled prices, courtesy E. Guevara H. \textit{et al} \cite{Esteban}]. Bottom: The largest eigenvalue $\lambda_1/N$ (circles) is a proxy for the average correlation (plain) [HY correlations for every pair and every bin of every day, then averaged over days for visual comfort and comparison with previous figure].} 
        \label{eigen}
\end{figure}

The largest eigenvalue $\lambda_1$ of the correlation matrix of stock returns, is well known to be associated with the ``market mode'', i.e. all stocks moving more or less in a synchronized manner. 
We show in the top panel of Fig.~\ref{eigen} the magnitude of $\lambda_1/N$ computed  from \eq{correlation} on 5-min data, as a function of the bin $k$. Interestingly, the average correlation clearly \textit{increases} as time elapses. 
As mentioned earlier, the quantity $\lambda_1/N$ captures the behavior of the average correlation between stocks, which can be seen in the bottom panel of Fig.~\ref{eigen}.

The evolution of the next six eigenvalues $\lambda_i(k)$, $i=2, \dots, 7$ is also shown in Fig.~\ref{eigen}. We see that the amplitudes of these {\it decrease} with time. It may be appropriate to quote the authors of Ref. \cite{Allez}: ``Although by construction the trace of the correlation matrix, and therefore the sum of all $N$ eigenvalues is constant (and equal to $N$), this decrease is not a trivial consequence of the increase of $\lambda_1$... What we see here is that as the day proceeds, more and more risk is carried 
by the market factor, while the amplitude of sectorial moves shrivels in relative terms (but remember that the correlation matrix is defined after normalising the returns by the local volatility, which increases in the last hours of the day).''

We also compute using \eq{HY} the cross-correlation matrices with tick-by tick data, for all 72 bins per day and 20 days in a month. The temporal evolution of the pairwise average correlation coefficients as a function of bins, for different days, and further averaged over all the days, are plotted below in \fg{average}.

\begin{figure} 
		\center
		\includegraphics[scale=0.35]{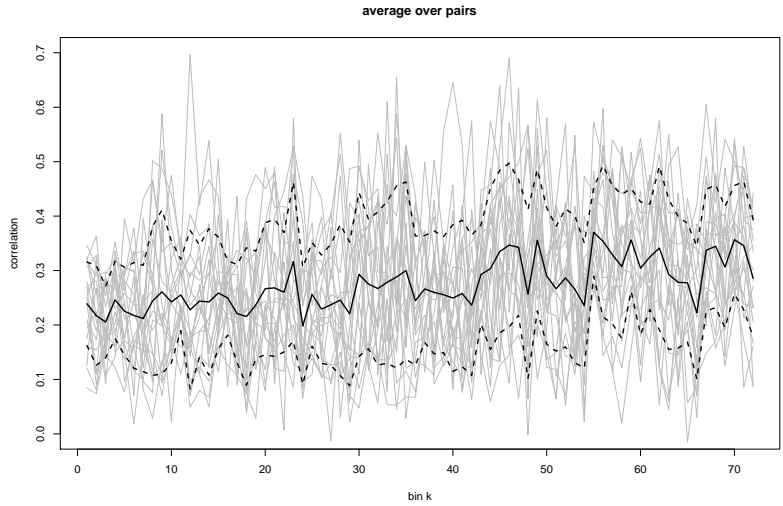}
		\caption{Plot of the (pairwise) average correlations as functions of bins $k$, for different days. Thick solid line: Plot of the average correlation coefficients, further averaged over all the days, which shows that the average correlation between stocks increases throughout the day. Thick dashed lines: Plots of the standard deviations on either side of the average correlation.} 
        \label{average}
\end{figure}

\subsection{MDS using intraday data}

In order to \textit{visually} capture the co-movement of stocks, we used the MDS plots of the 40 stocks of the CAC40 index (see list of CAC40 stocks in Table \ref{rics}), for the period of March 2011. We used 30 minute bins to compute the correlations, using the Hayashi-Yoshida estimator. We used the period 10h00-16h00 CET, so as to get 12 bins per day for the 22 days. Using the correlation matrices as input, we made the distance transformations (using \eq{distance}) to produce the distance matrices. These distance matrices were then used as inputs to the standard MDS function in \texttt{MATLAB}.  
We used the method of simulated annealing to optimize the cost function of a particular bin. The first bin starts with an initial set of coordinates chosen at random; for the following bins, we used the final results of the previous bins as the initial states\footnote{This is to avoid too drastic a change in the MDS plots from one bin to another, keeping in mind that the vectors $x_i$ are \textit{not unique}-- with the Euclidean metric, they may be arbitrarily \textit{translated} and \textit{rotated}.}.
The output of the MDS were the coordinates, which were plotted as the MDS maps. The coordinates were plotted in a manner such that the centroid of the map coincided with the origin $(0,0)$. We then computed the mean distance of all the coordinates from the centre, and plotted this measure as a  functon of time. 

During the course of any day, since for every bin the correlation matrix changes, the MDS map also changes. Just as it is interesting  to study how the average correlation between the stocks varies during the day, we thought it would be also interesting to study how the MDS map evolves ``on an average'' during the day. We had two choices: (i) Run the MDS algorithm for every bin for 22 days, and take the average of the coordinates over all the 22 maps, and plot this map for every bin. (ii) Take the average of the correlations over the 22 days for each bin, and plot a single MDS map for every bin. We executed both, to see the variations. In choice (i), for every bin $k$ we take an average of the coordinates generated by the 22 MDS runs (for different days) and plot them stock by stock. Some stocks fluctuate a lot on a day to day basis, in the same time bin; others fluctuate less. On the whole we expected to see the average structure (clustering) of the market. In choice (ii), we expected to see less structure, since when we take the average of correlations over all 22 days, and then run the MDS once for every bin, the variances in the correlations disappear and so the MDS plots look more uniform.

\subsubsection{Averaged (over days) coordinates in different bins}

We took the average of the coordinates (output of the MDS) of each company over all 22 days, for a particular bin. We then plotted the MDS maps using these averaged coordinates for the different bins to see the evolution during the day, as shown in \fg{MDS_avg_coordinates1} (for first six bins) and \fg{MDS_avg_coordinates2} (for last six bins). We find that there is some structure, and particular companies always stay together in a cluster or a group.
\begin{figure}[H]
\begin{center}
\includegraphics[width=0.49\linewidth]{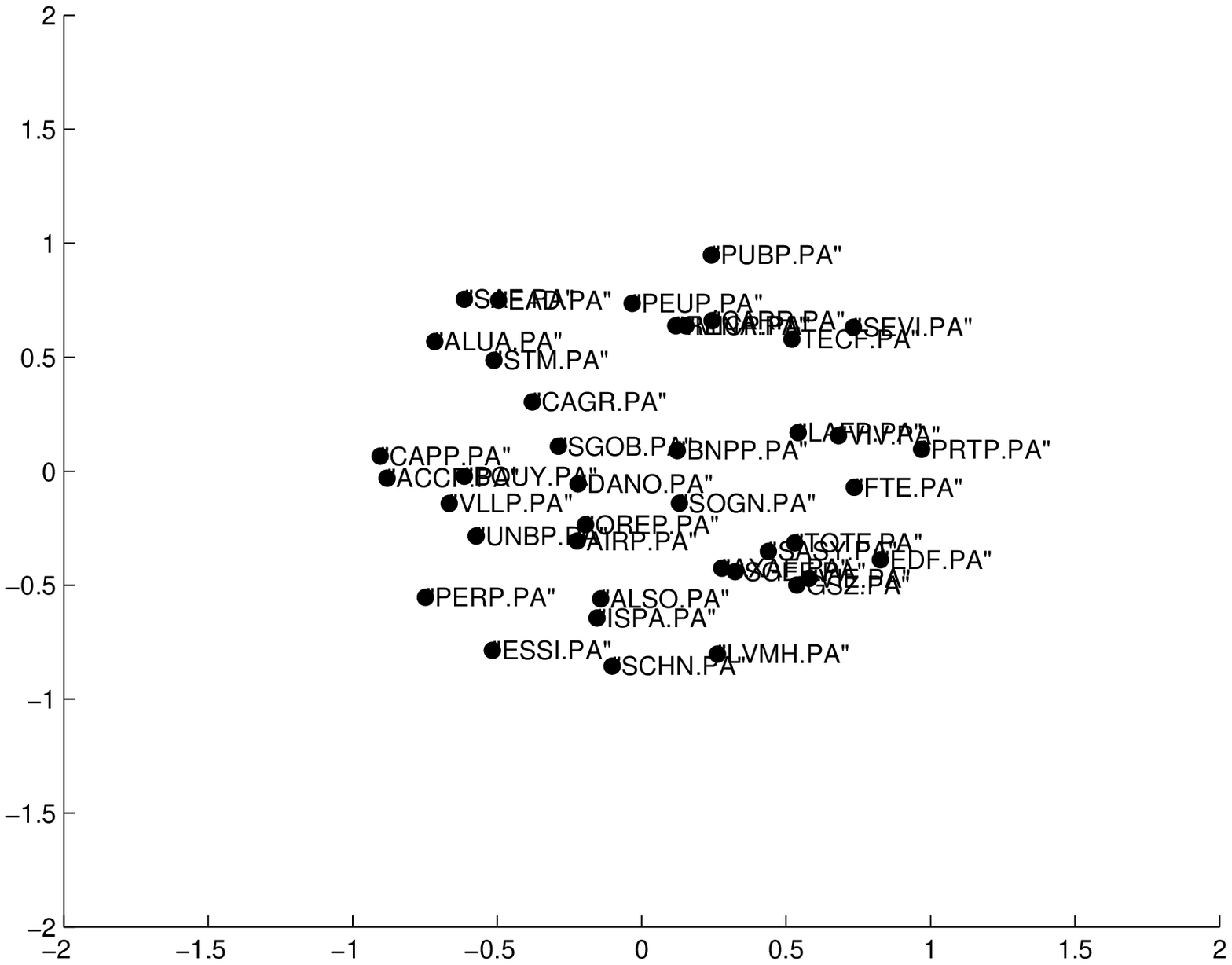}
\includegraphics[width=0.49\linewidth]{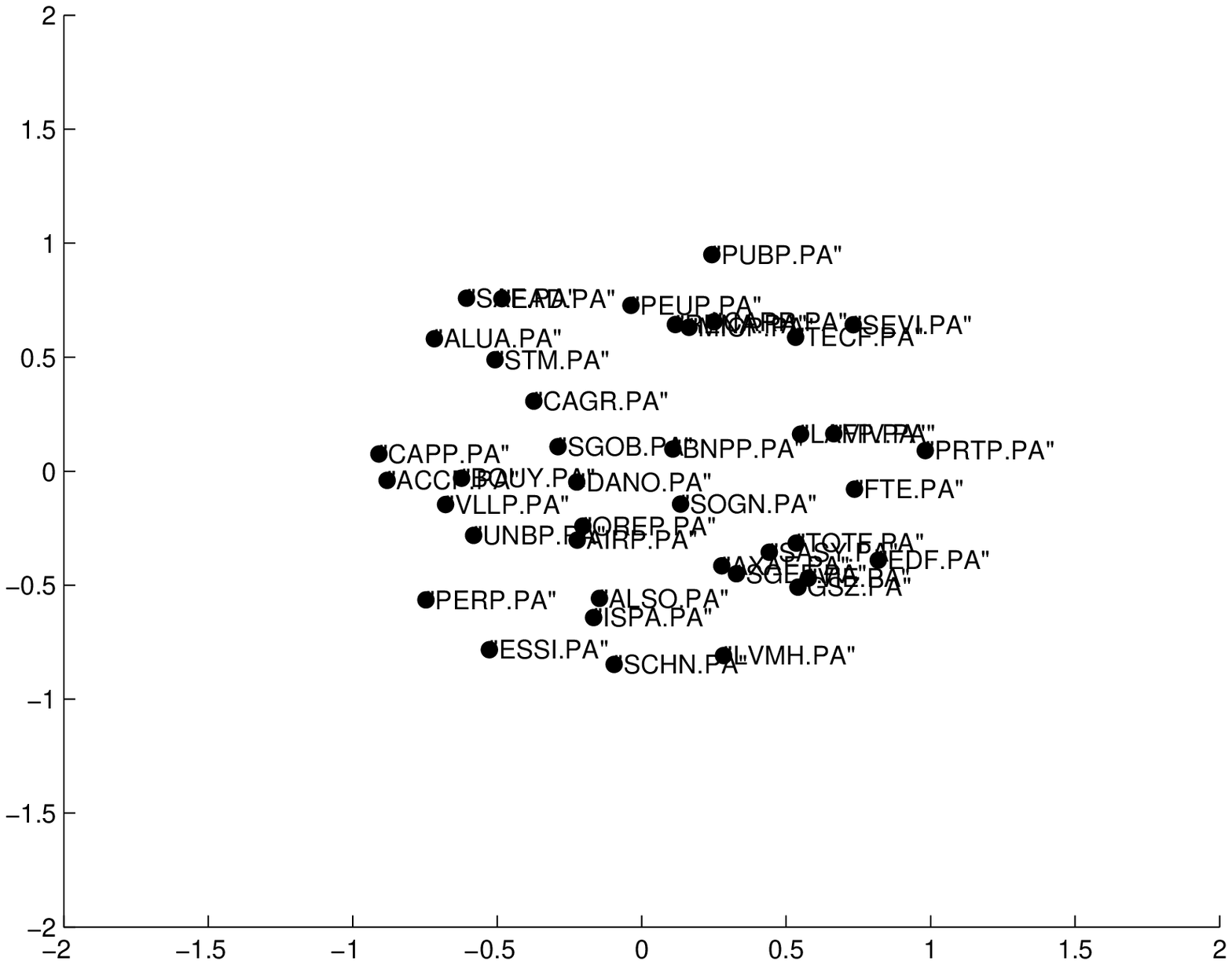}\\
\includegraphics[width=0.49\linewidth]{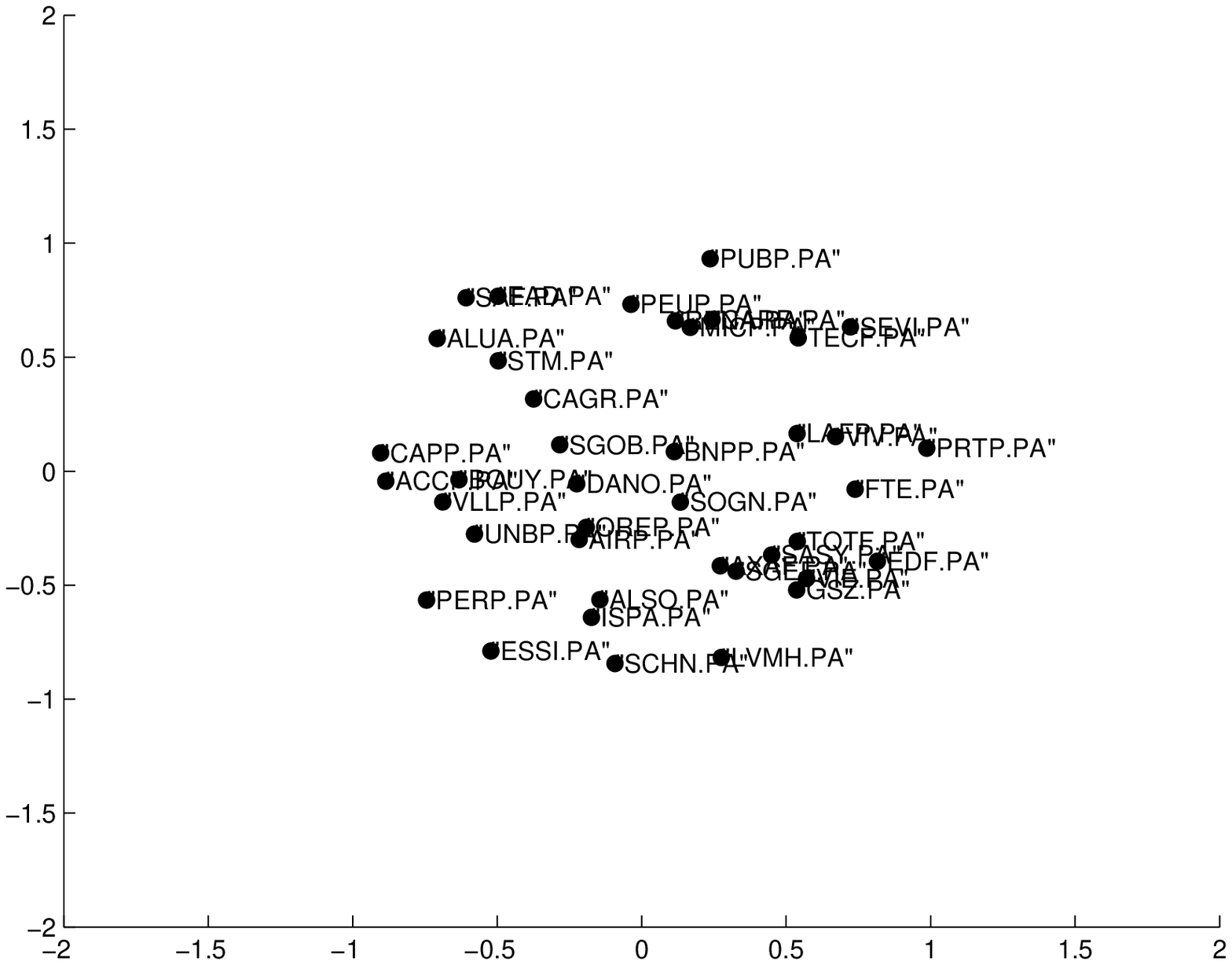}
\includegraphics[width=0.49\linewidth]{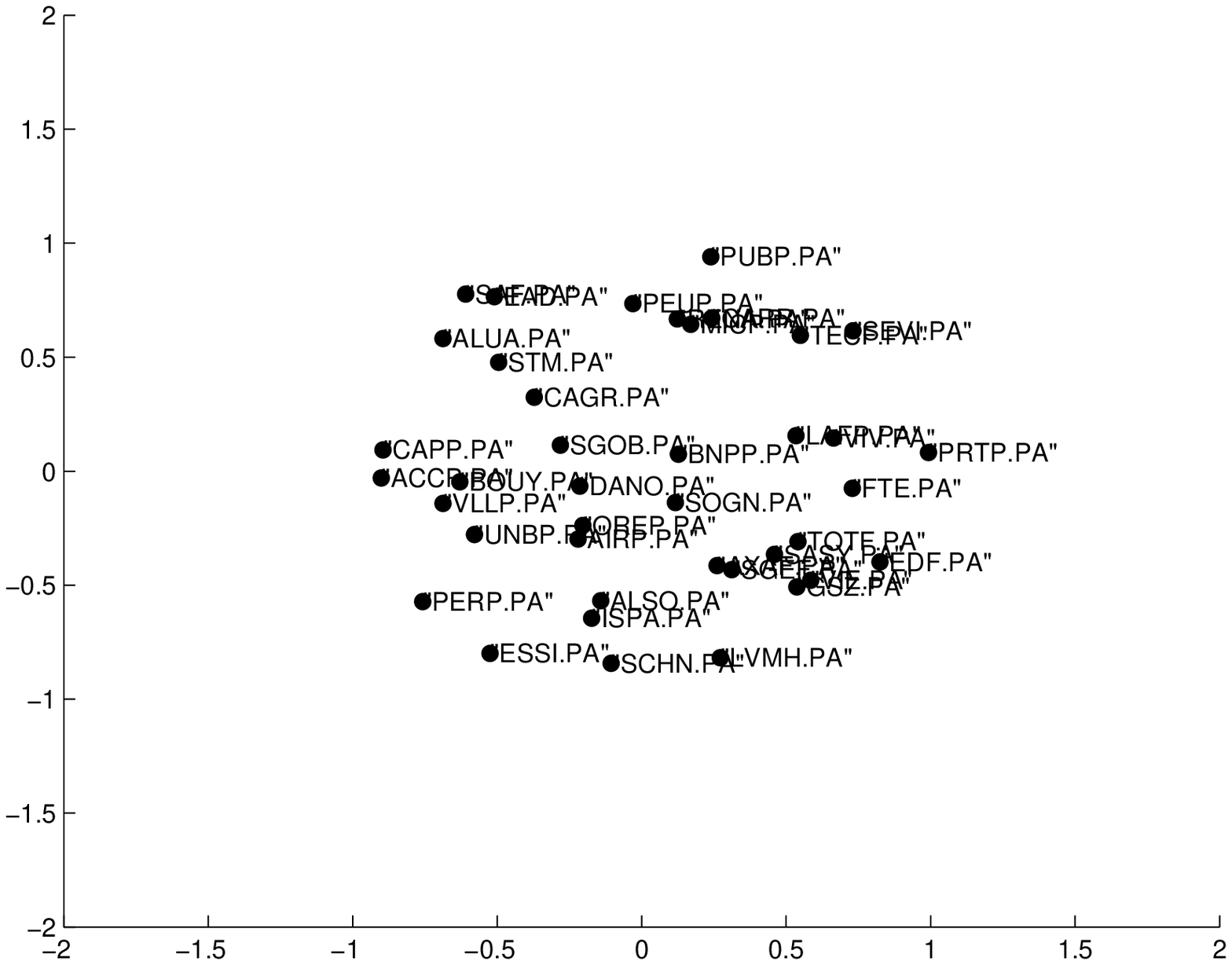}\\
\includegraphics[width=0.49\linewidth]{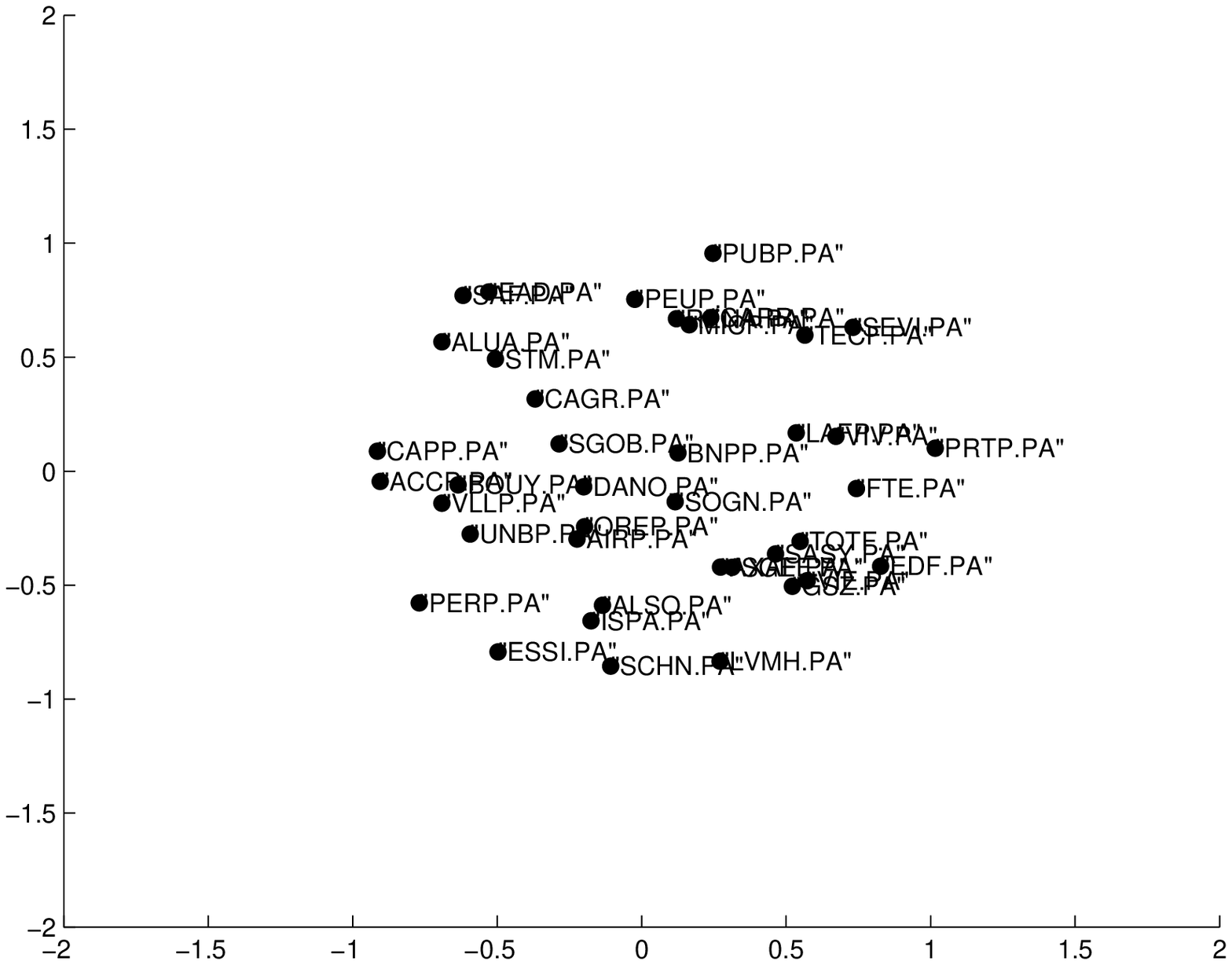}
\includegraphics[width=0.49\linewidth]{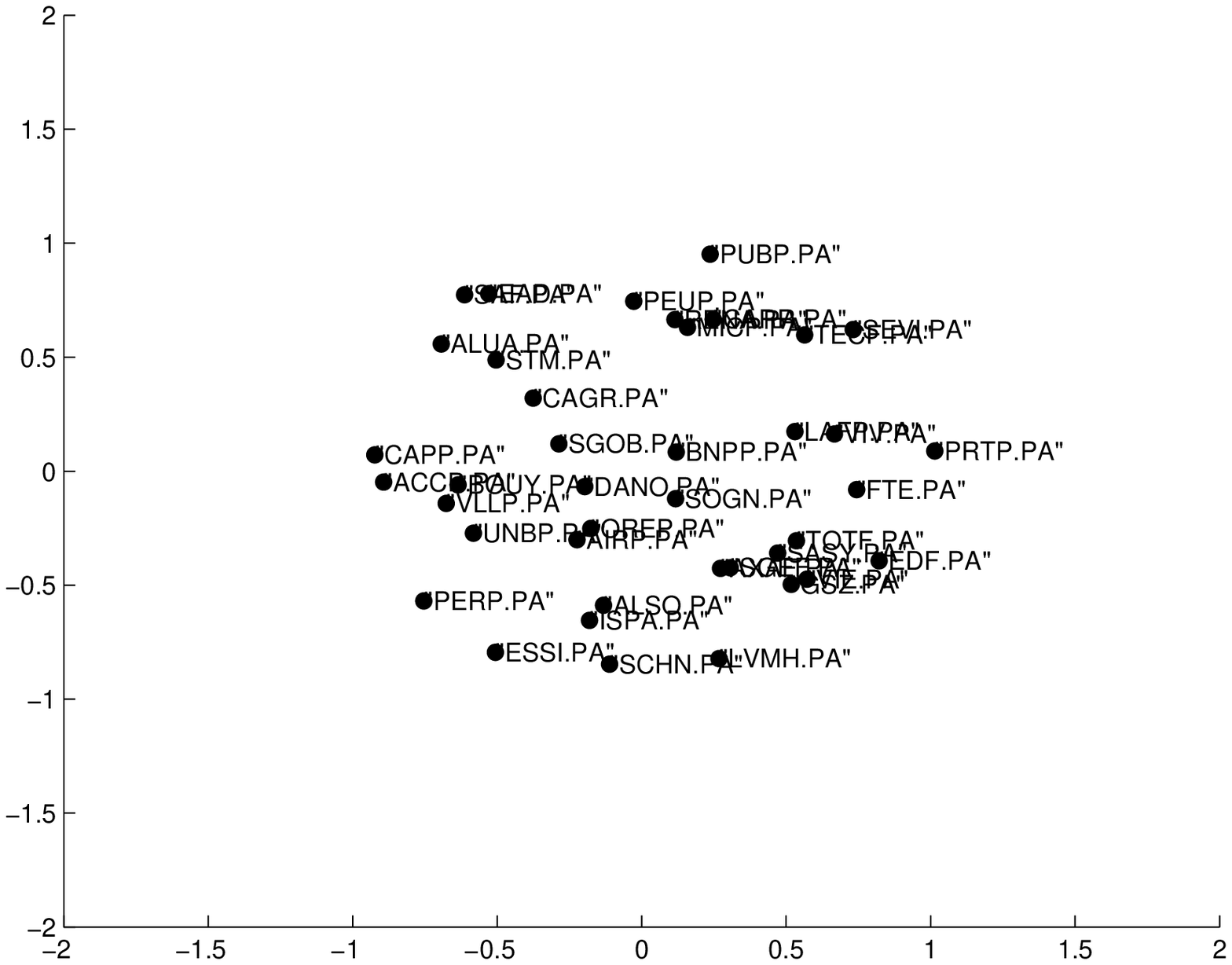}
\end{center}
\caption{MDS plots for bins 1-6. Each point on a plot represents a stock (see list of CAC40 stocks in Table \ref{rics}), designated by two coordinates ($x_i,y_i$), $i=1, \dots, N$. We took the average of the coordinates (output of the MDS) of each company over all 22 days, for a particular bin. We then plotted the MDS maps using these averaged coordinates for the different bins to see the evolution during the day. }
\label{MDS_avg_coordinates1}
\end{figure}

\begin{figure}[H]
\begin{center}
\includegraphics[width=0.49\linewidth]{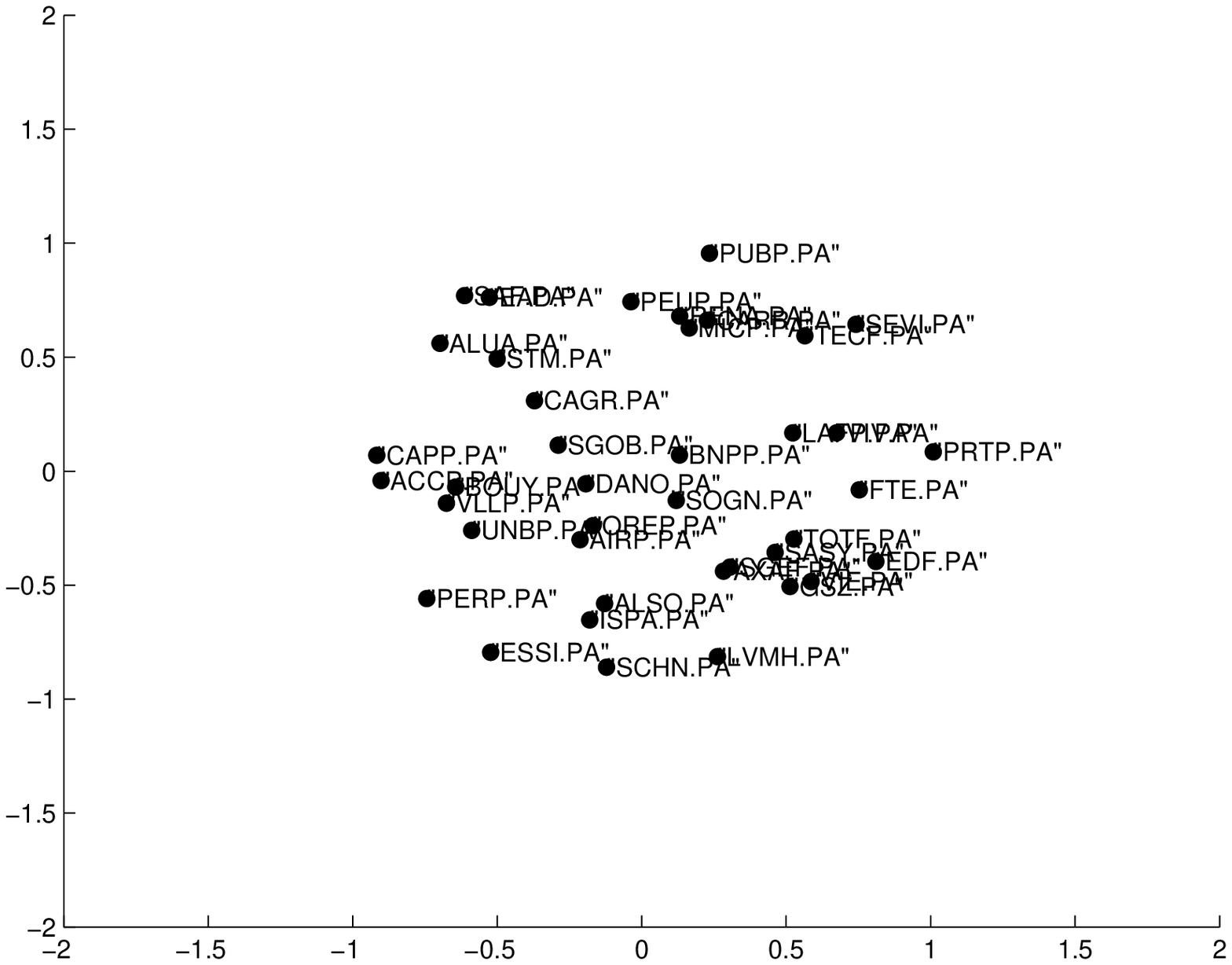}
\includegraphics[width=0.49\linewidth]{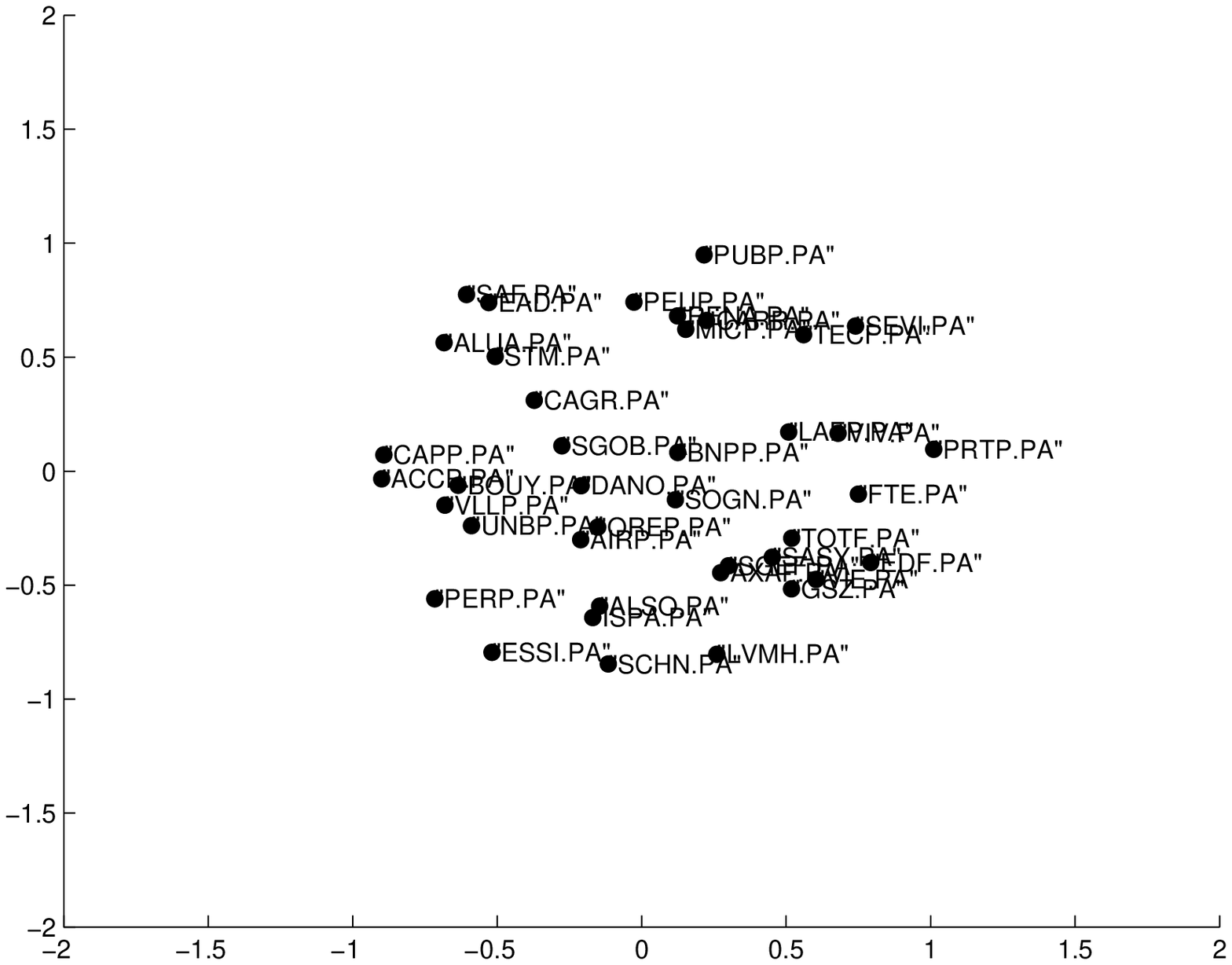}\\
\includegraphics[width=0.49\linewidth]{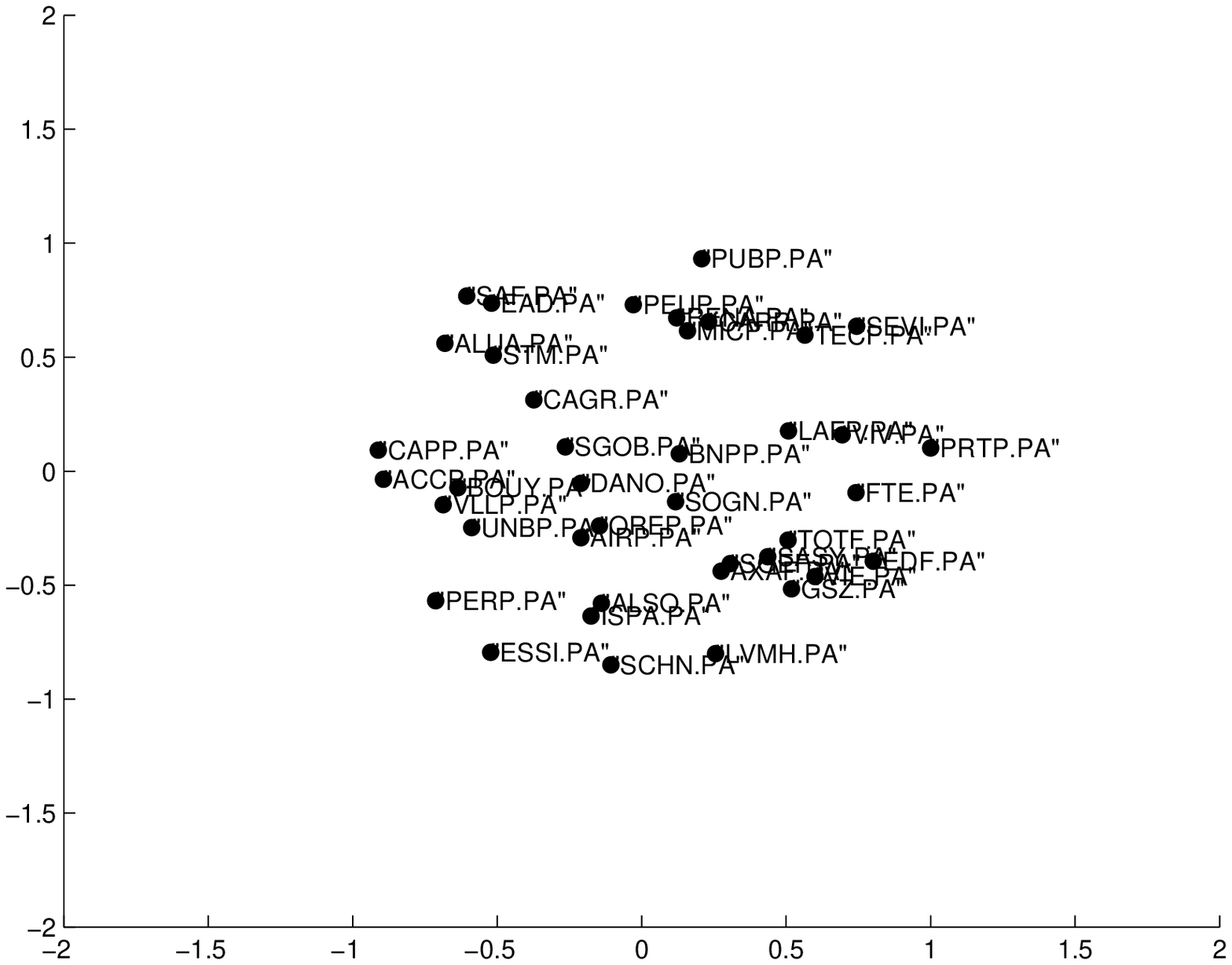}
\includegraphics[width=0.49\linewidth]{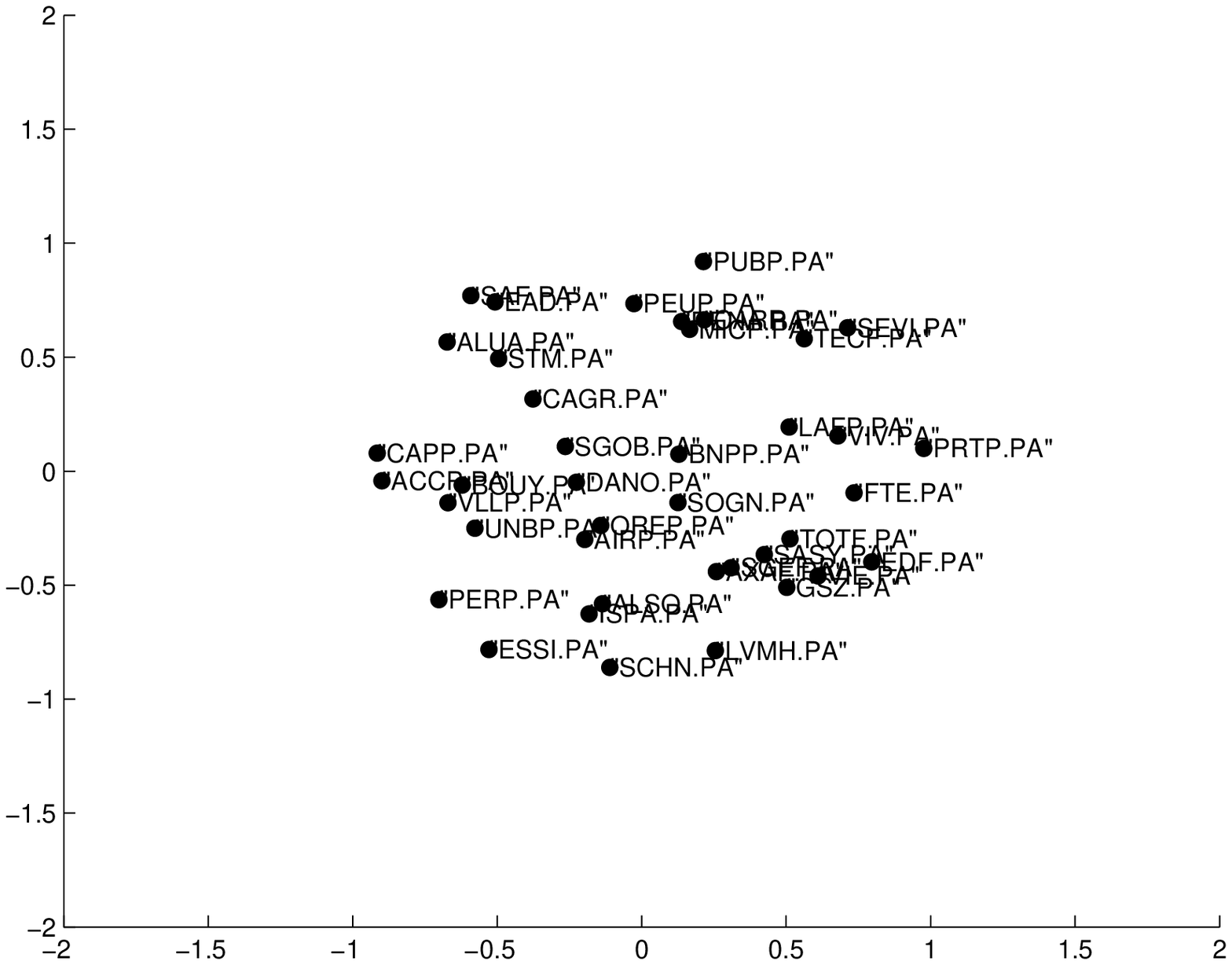}\\
\includegraphics[width=0.49\linewidth]{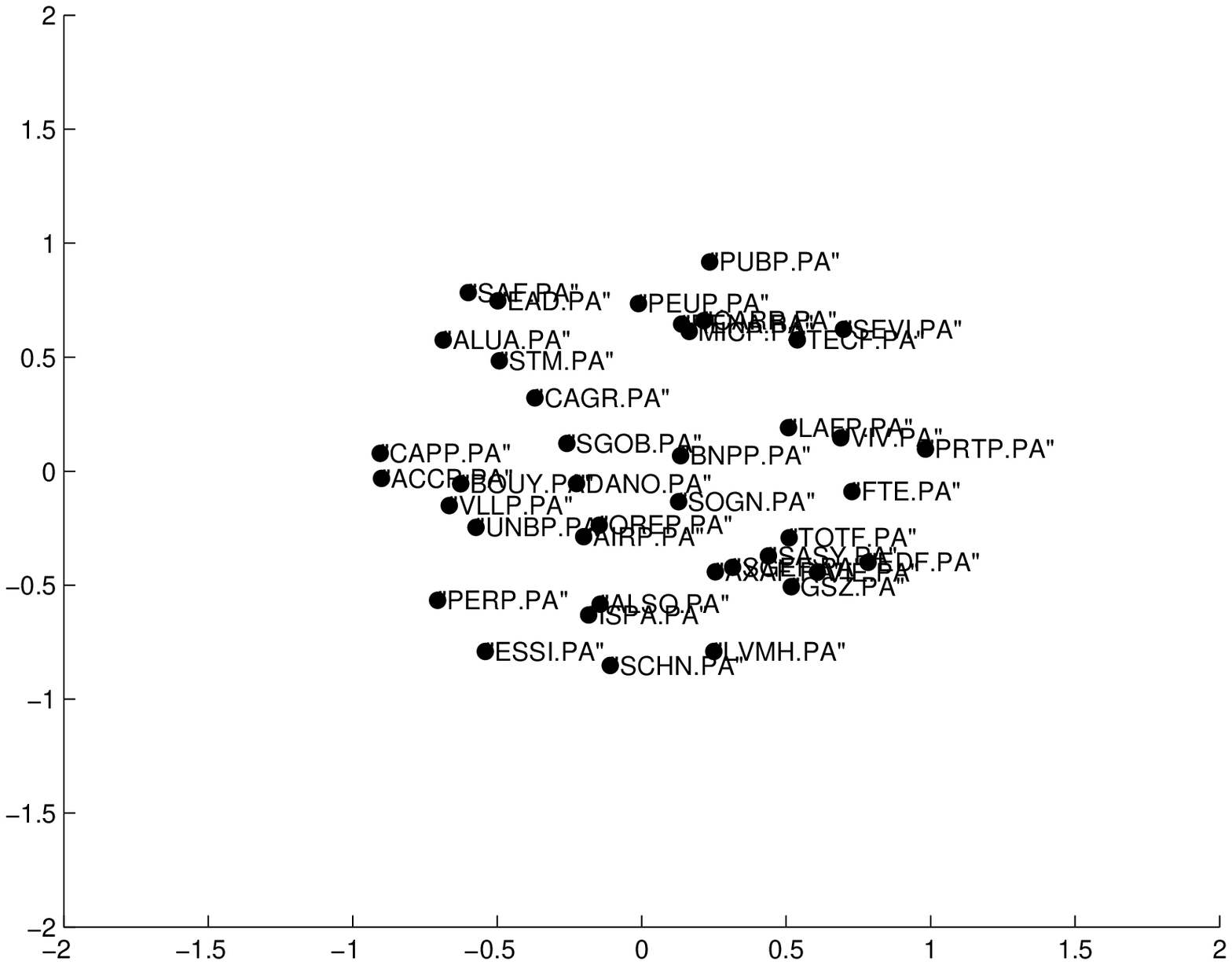}
\includegraphics[width=0.49\linewidth]{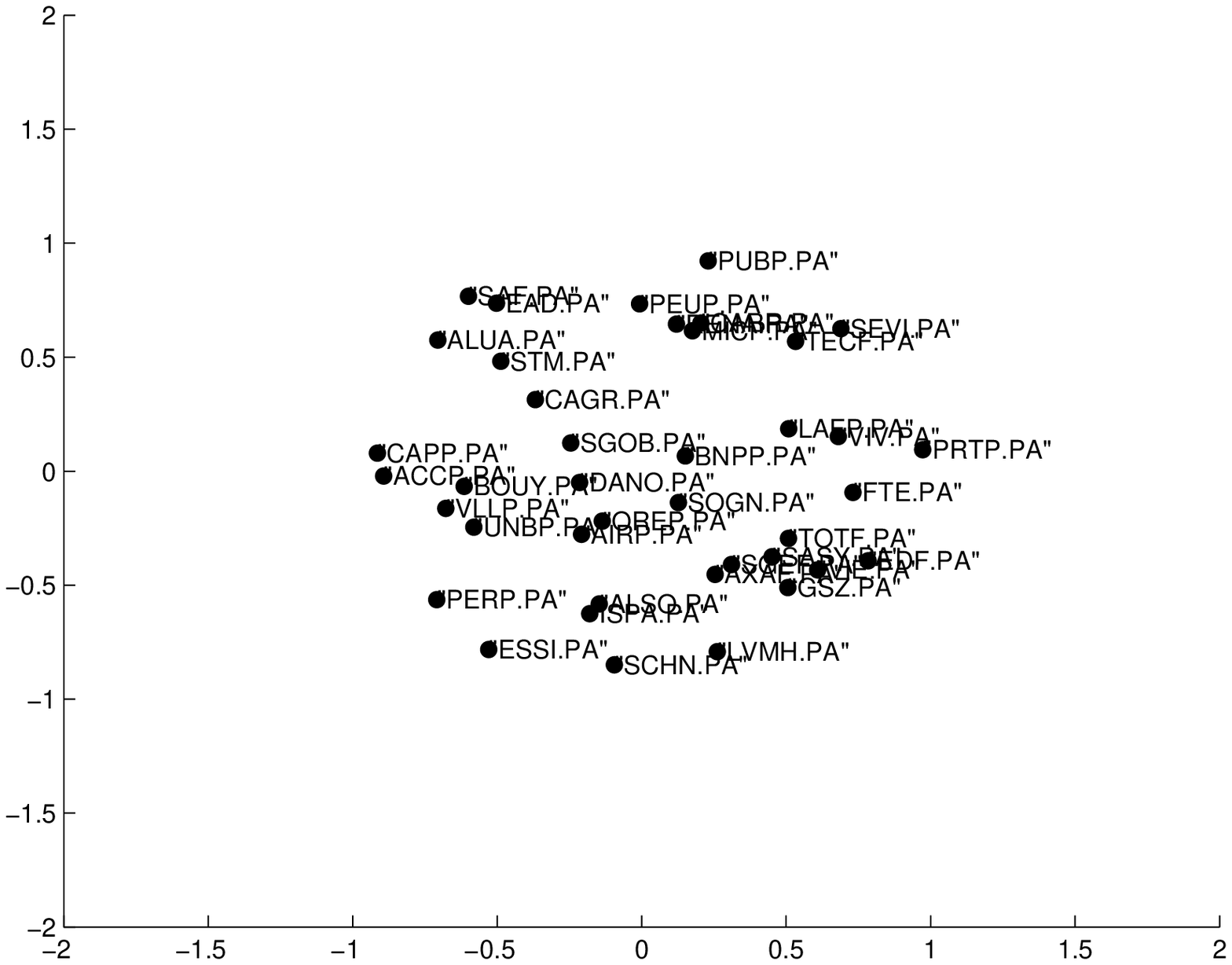}
\end{center}
\caption{MDS plots  for bins 7-12. Each point on a plot represents a stock (see list of CAC40 stocks in Table \ref{rics}), designated by two coordinates ($x_i,y_i$), $i=1, \dots, N$. We took the average of the coordinates (output of the MDS) of each company over all 22 days, for a particular bin. We then plotted the MDS maps using these averaged coordinates for the different bins to see the evolution during the day. }
\label{MDS_avg_coordinates2}
\end{figure}

\subsubsection{Averaged (over days) correlations in different bins}

We also took the average of the correlation coefficients for each pair over all 22 days, and then used them to generate the MDS plot for a particular bin. We then plotted the MDS maps for the different bins to see the evolution during the day, as shown in \fg{MDS_avg_correlations1} (for first six bins) and \fg{MDS_avg_correlations2} (for last six bins). We find that there is less structure than the previous plots (as average of correlations ``smoothen out'' the dissimilarities). The structures of the maps and positions of the companies do not change drastically during the course of the day.

\begin{figure}[H]
\begin{center}
\includegraphics[width=0.49\linewidth]{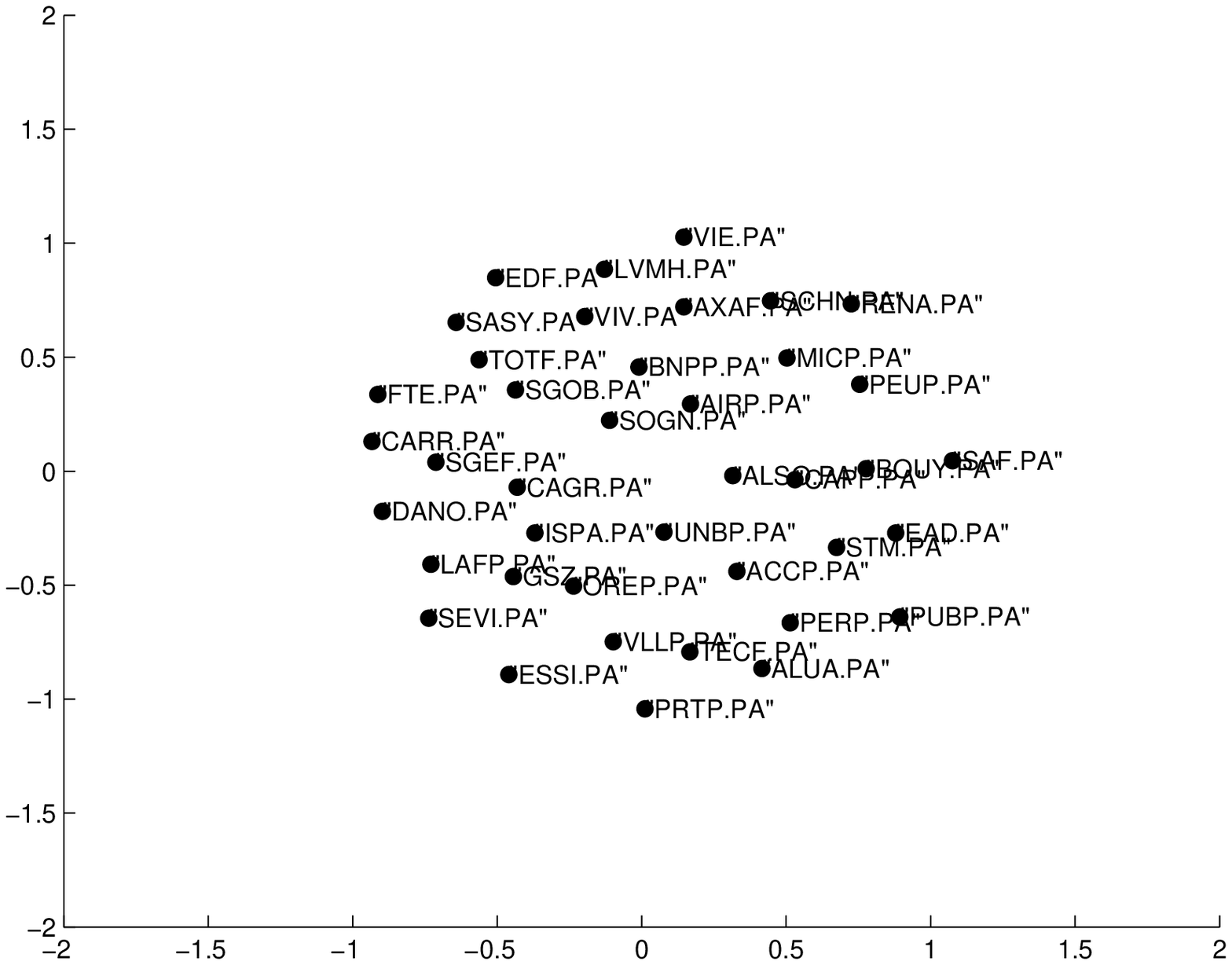}
\includegraphics[width=0.49\linewidth]{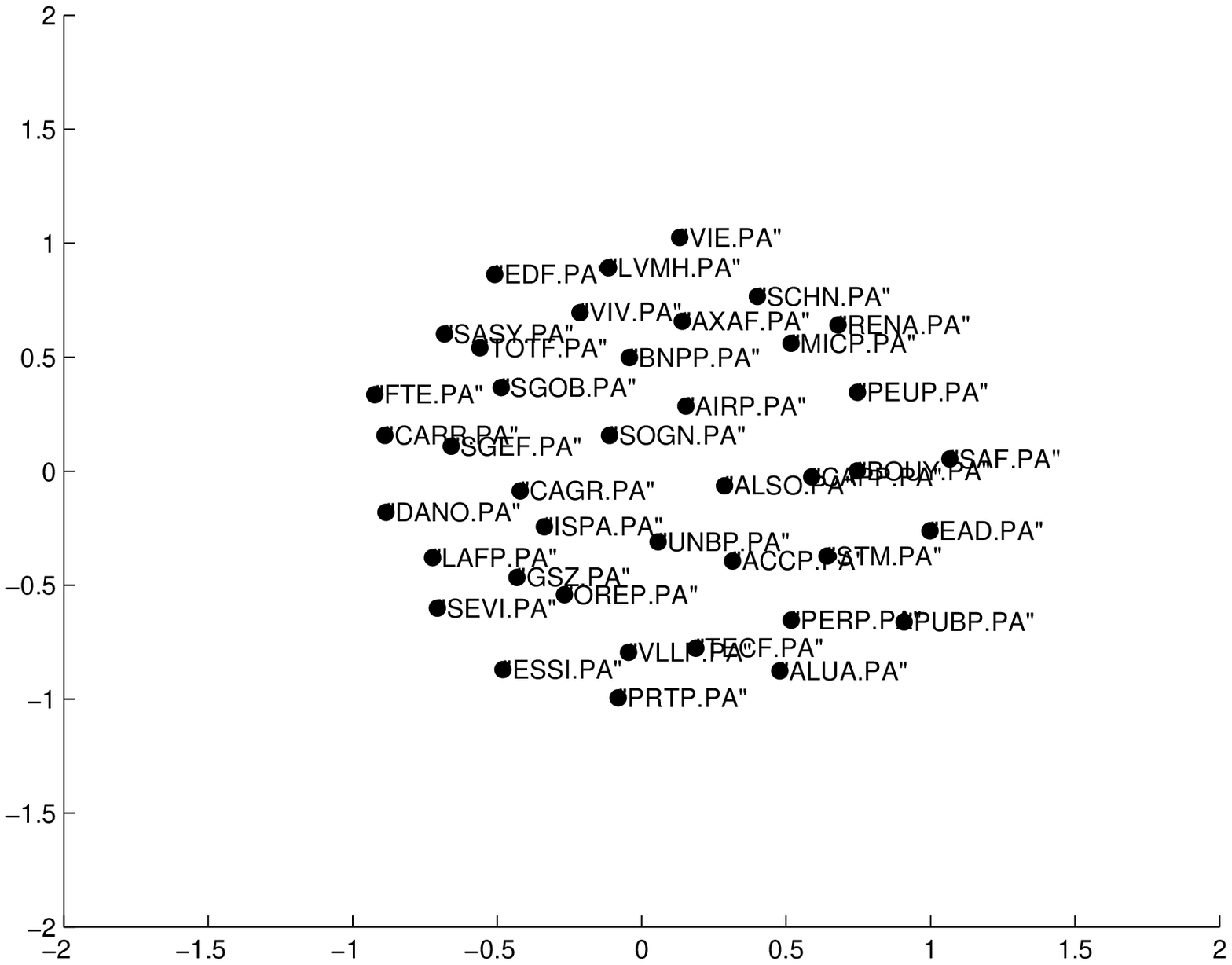}\\
\includegraphics[width=0.49\linewidth]{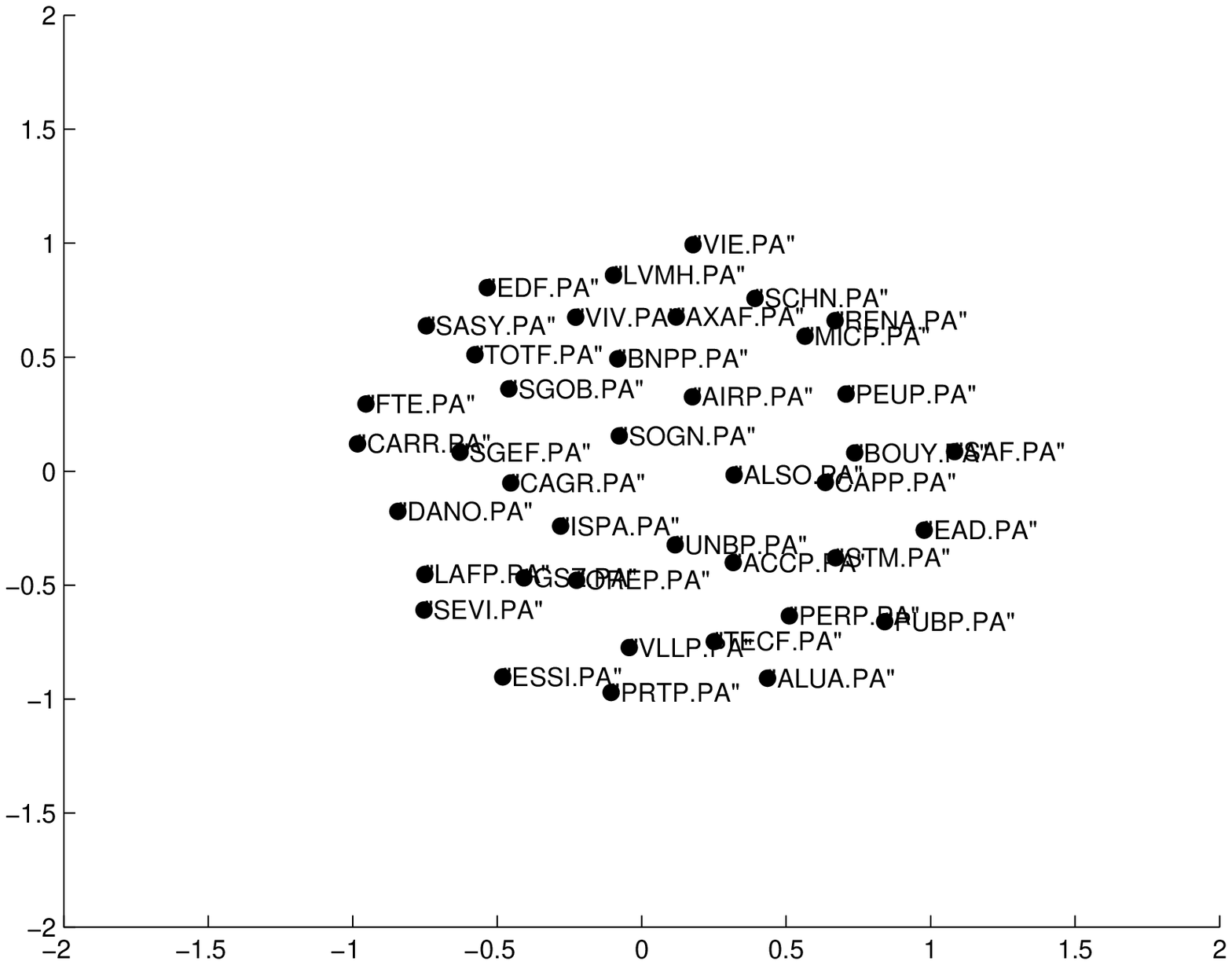}
\includegraphics[width=0.49\linewidth]{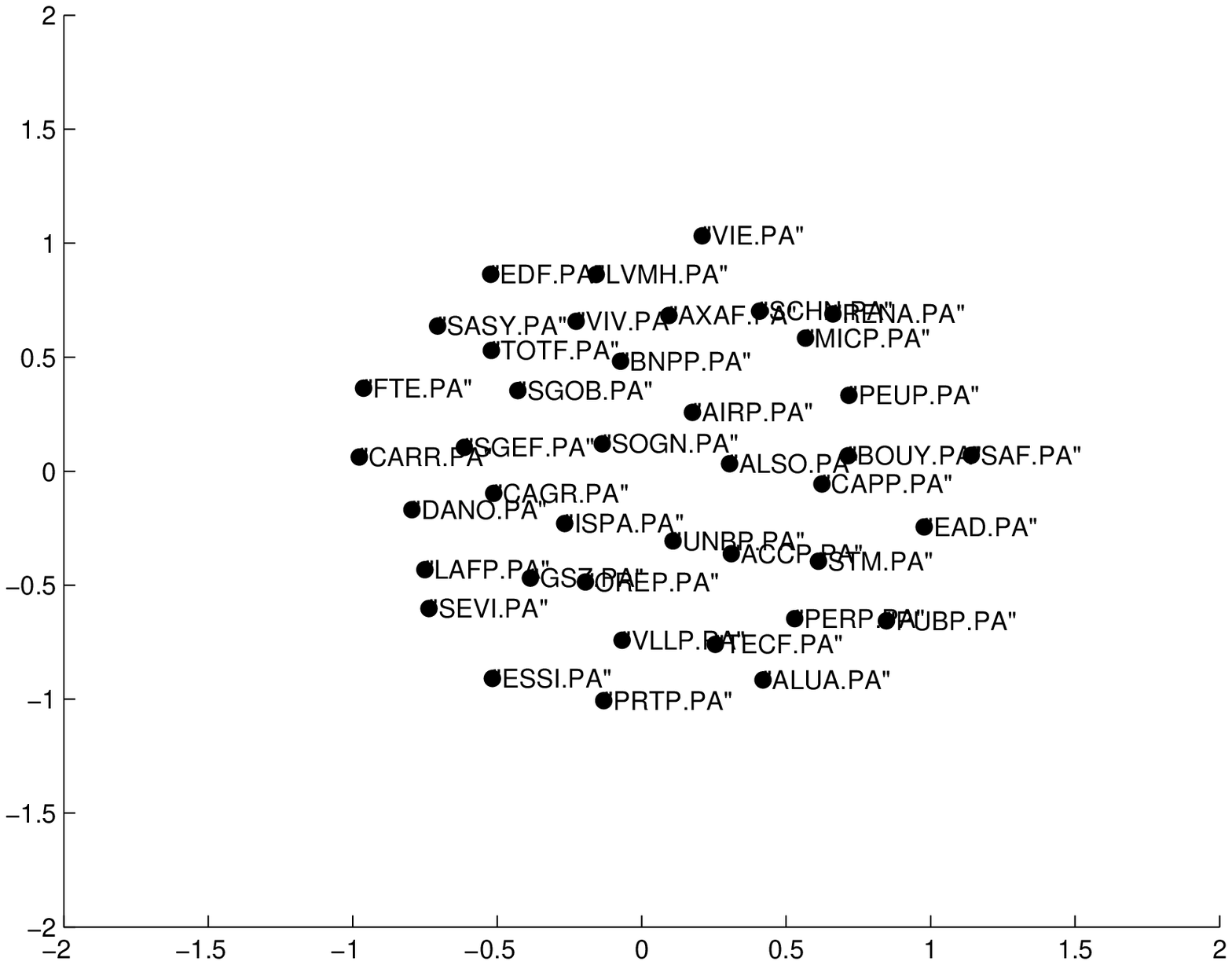}\\
\includegraphics[width=0.49\linewidth]{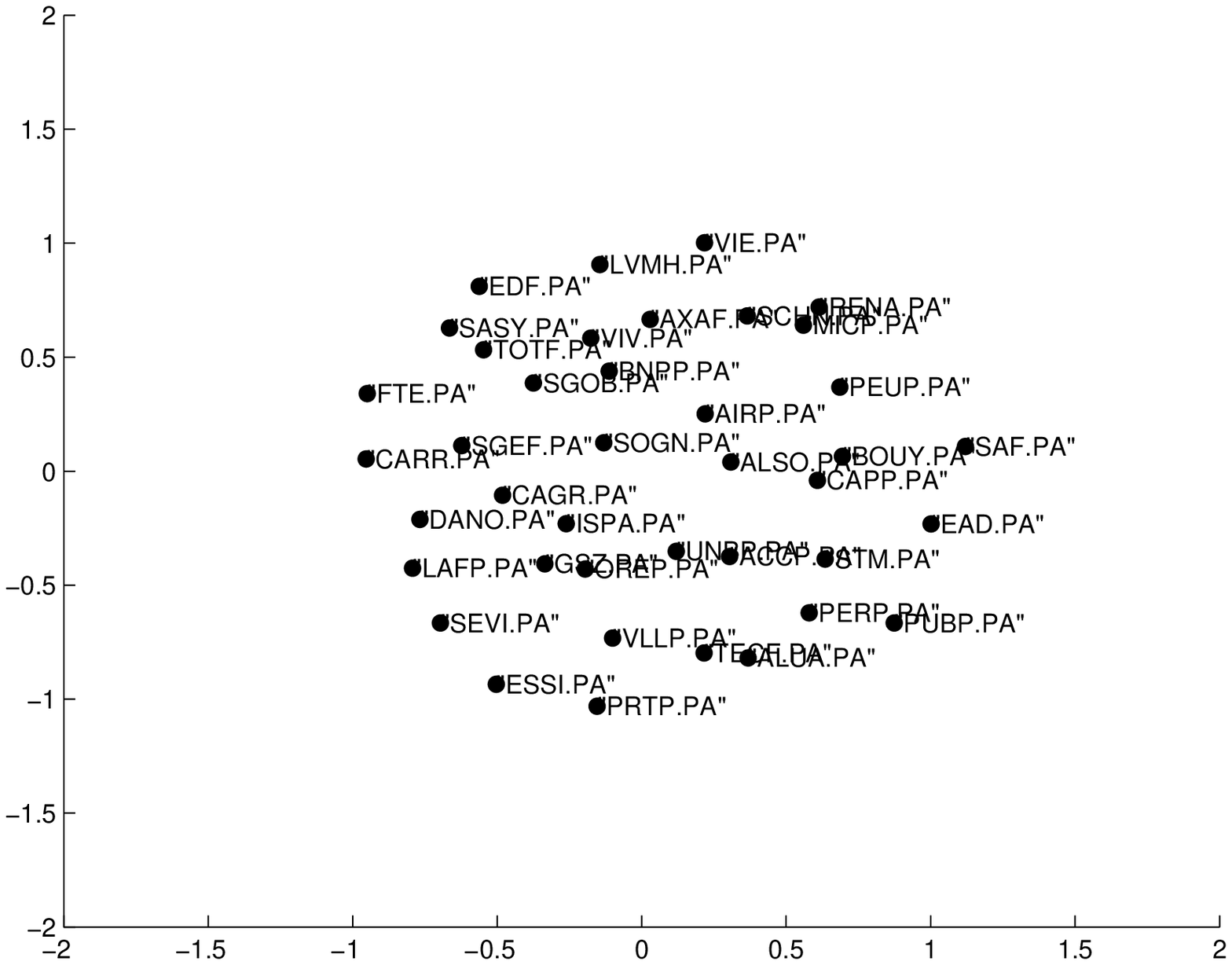}
\includegraphics[width=0.49\linewidth]{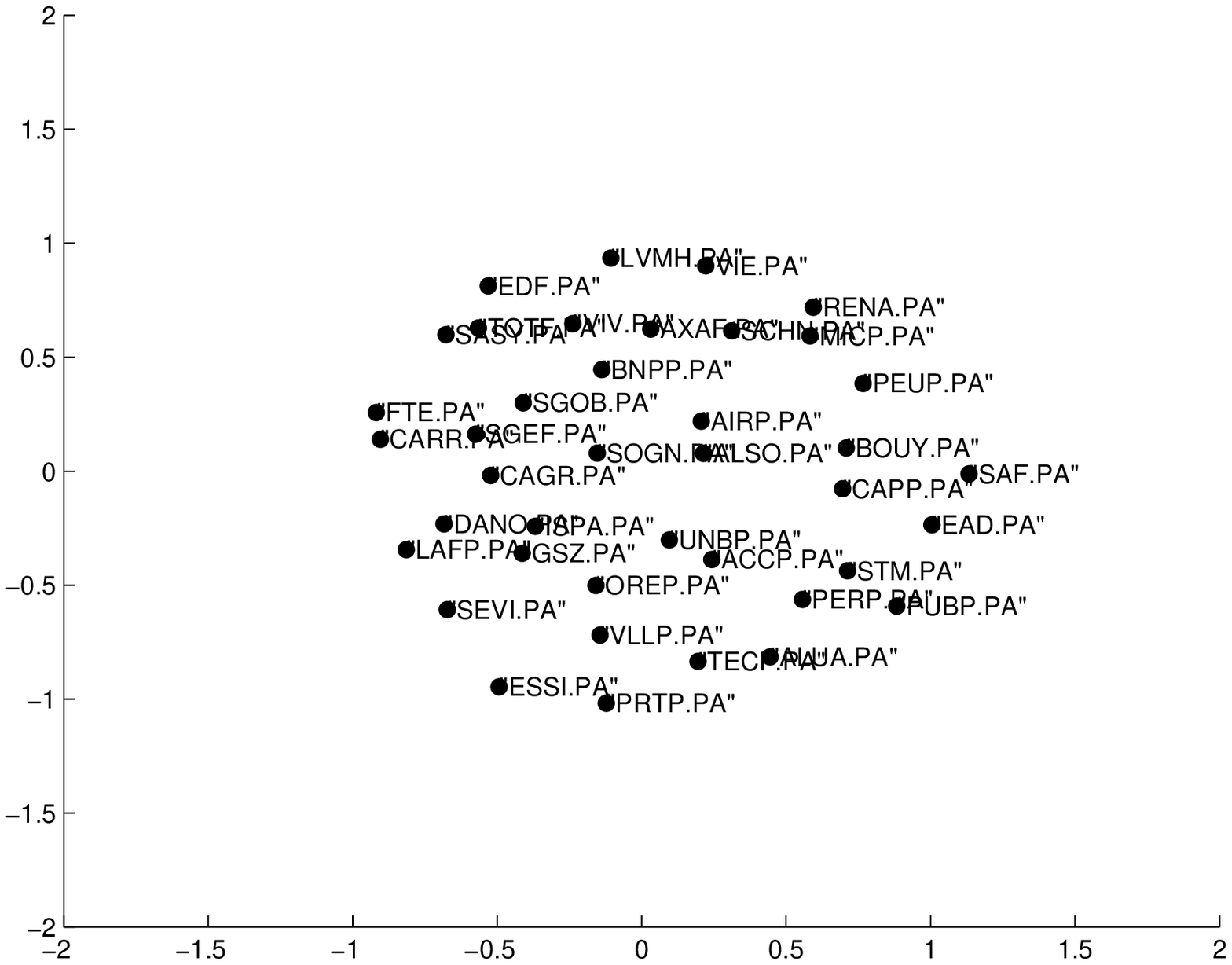}
\end{center}
\caption{MDS plots  for bins 1-6. Each point on a plot represents a stock (see list of CAC40 stocks in Table \ref{rics}), designated by two coordinates ($x_i,y_i$), $i=1, \dots, N$. We took the average of the correlation coefficients for each pair over all 22 days, and then used them to generate the MDS plot for a particular bin. We then plotted the MDS maps for the different bins to see the evolution during the day.}
\label{MDS_avg_correlations1}
\end{figure}

\begin{figure}[H]
\begin{center}
\includegraphics[width=0.49\linewidth]{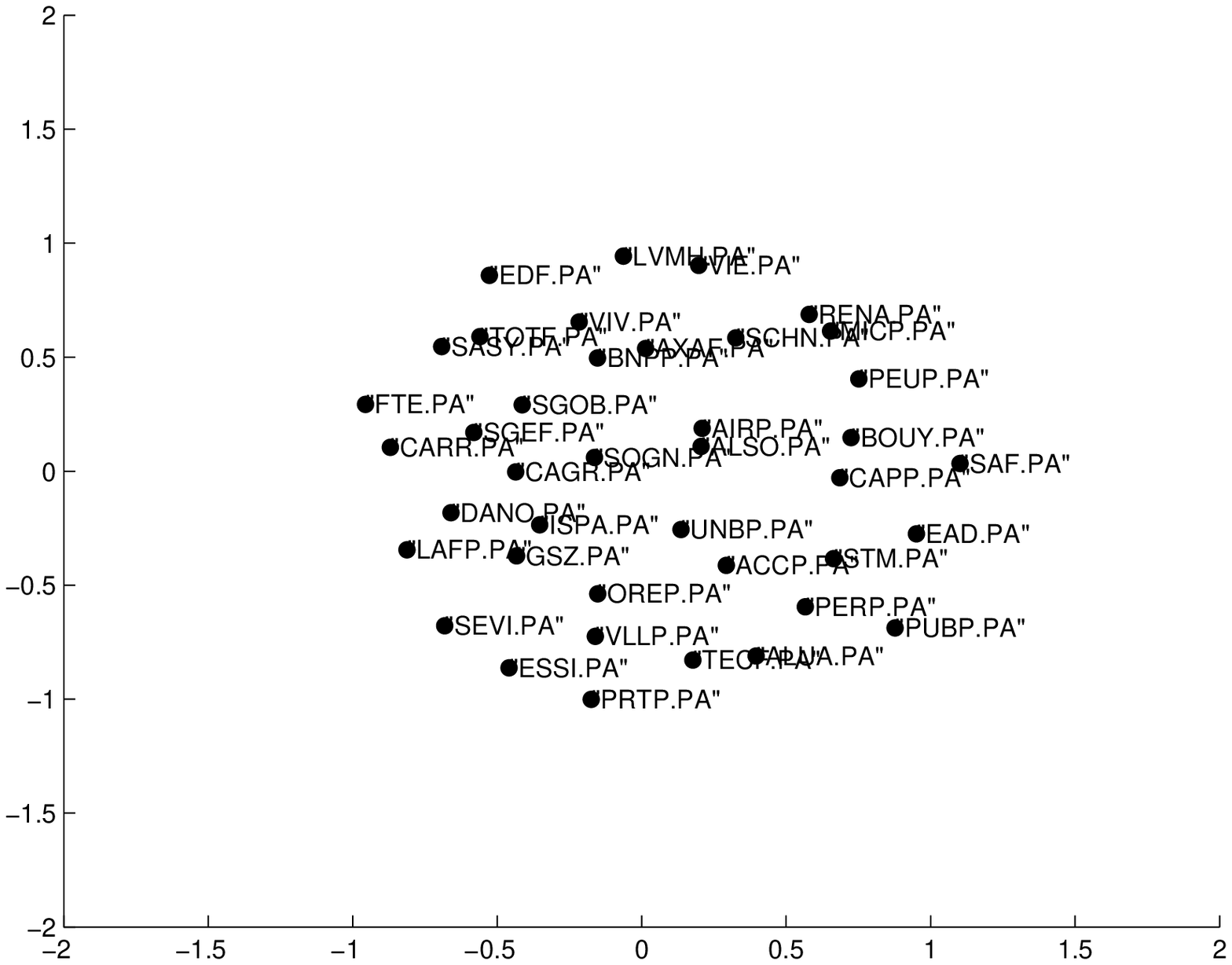}
\includegraphics[width=0.49\linewidth]{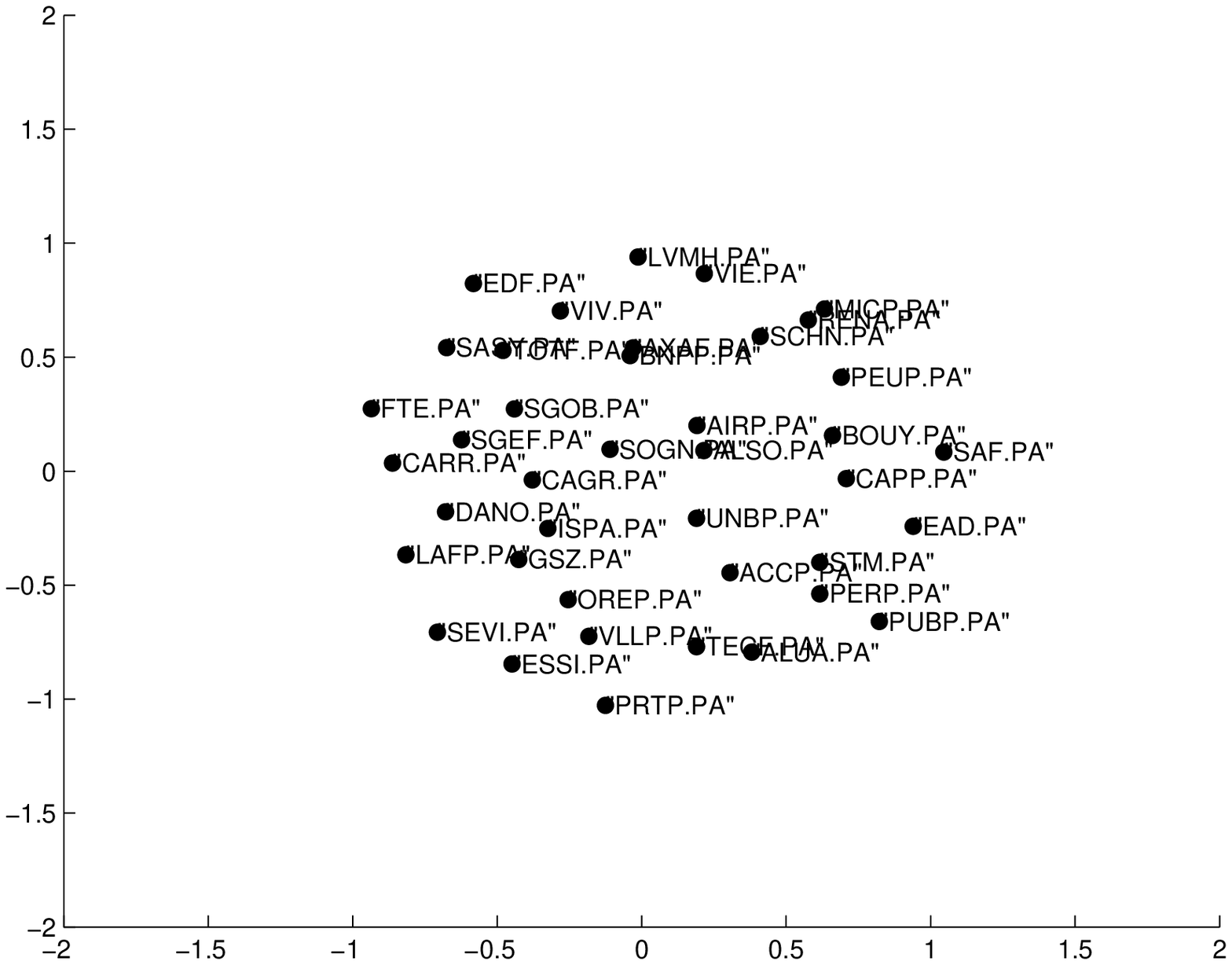}\\
\includegraphics[width=0.49\linewidth]{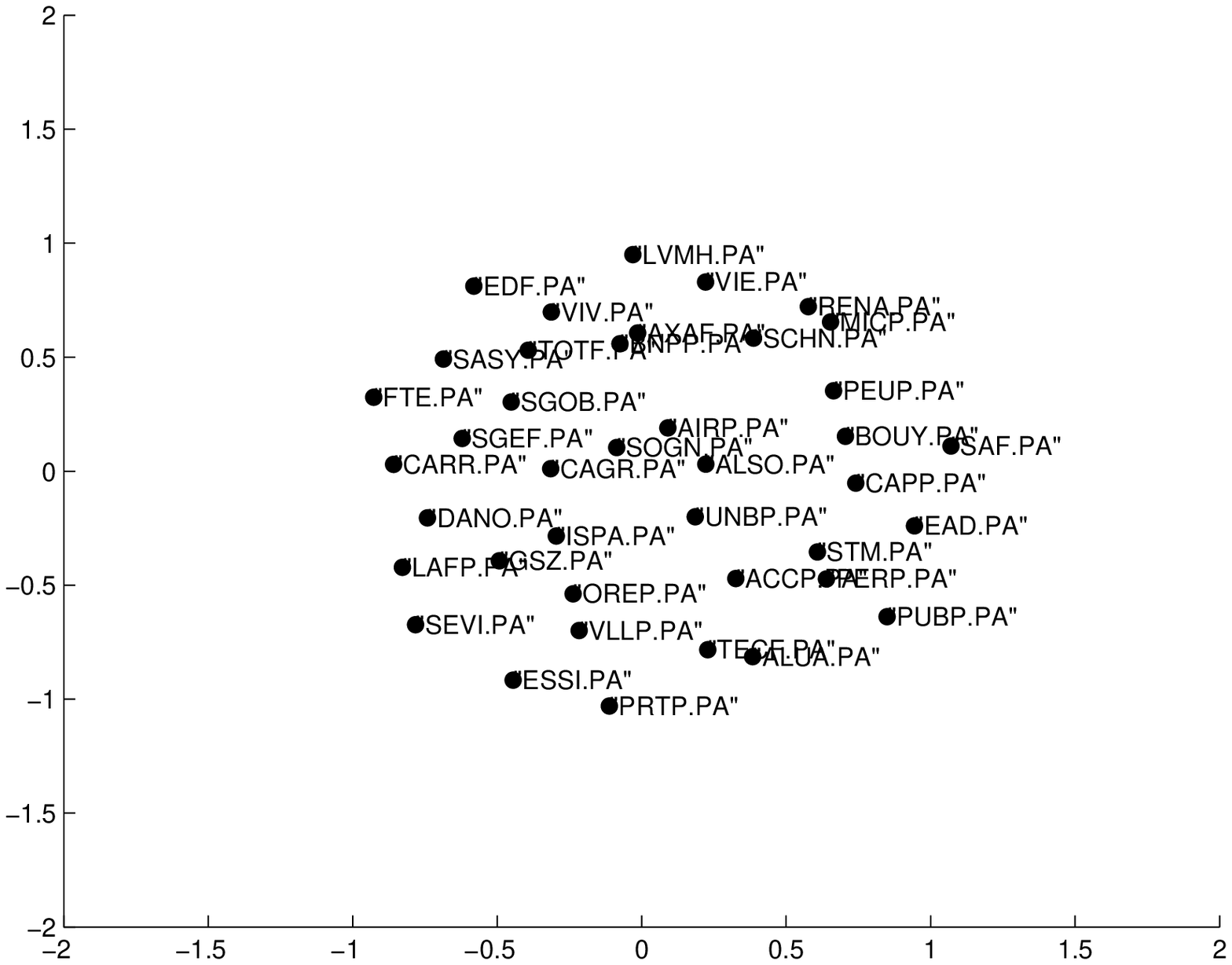}
\includegraphics[width=0.49\linewidth]{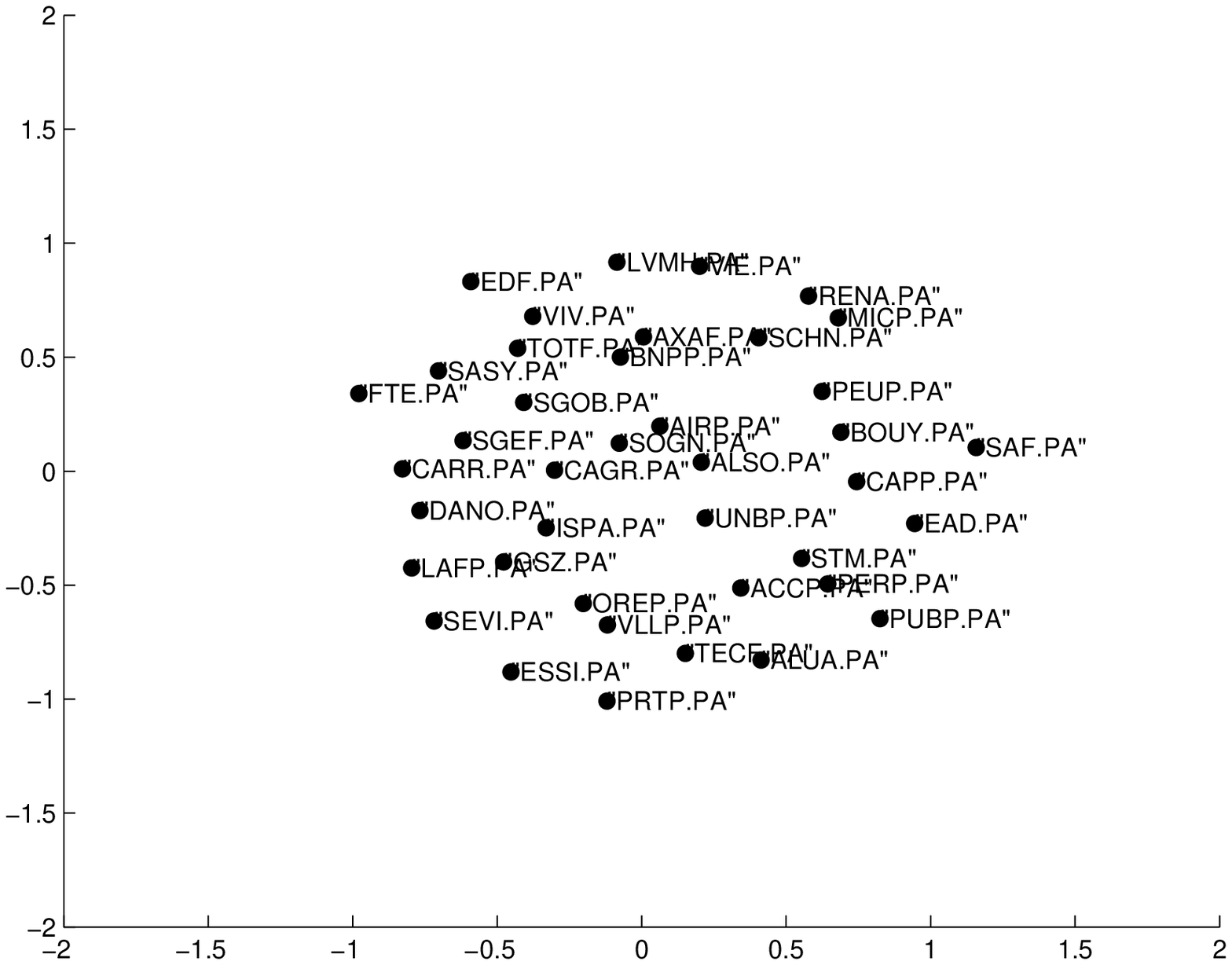}\\
\includegraphics[width=0.49\linewidth]{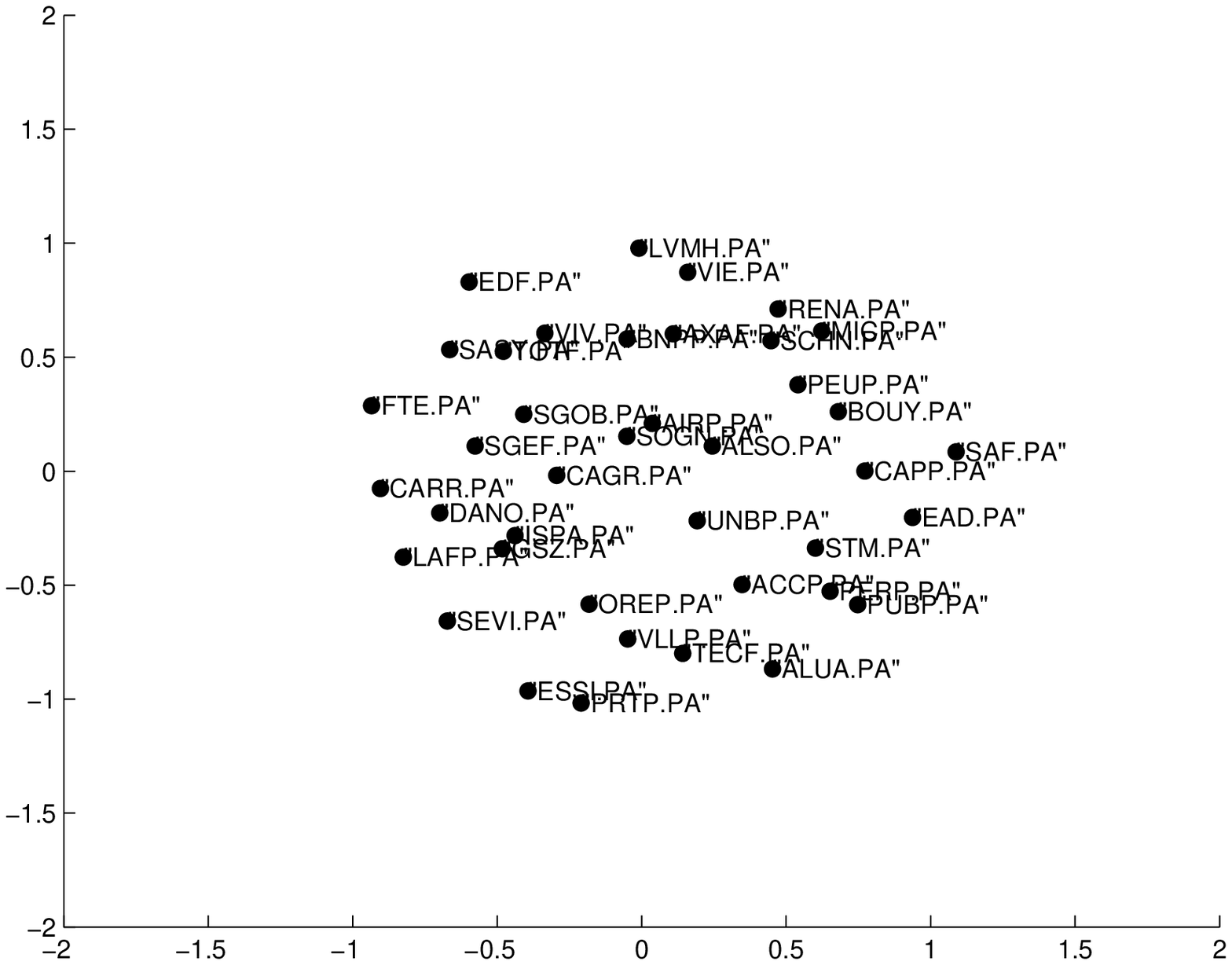}
\includegraphics[width=0.49\linewidth]{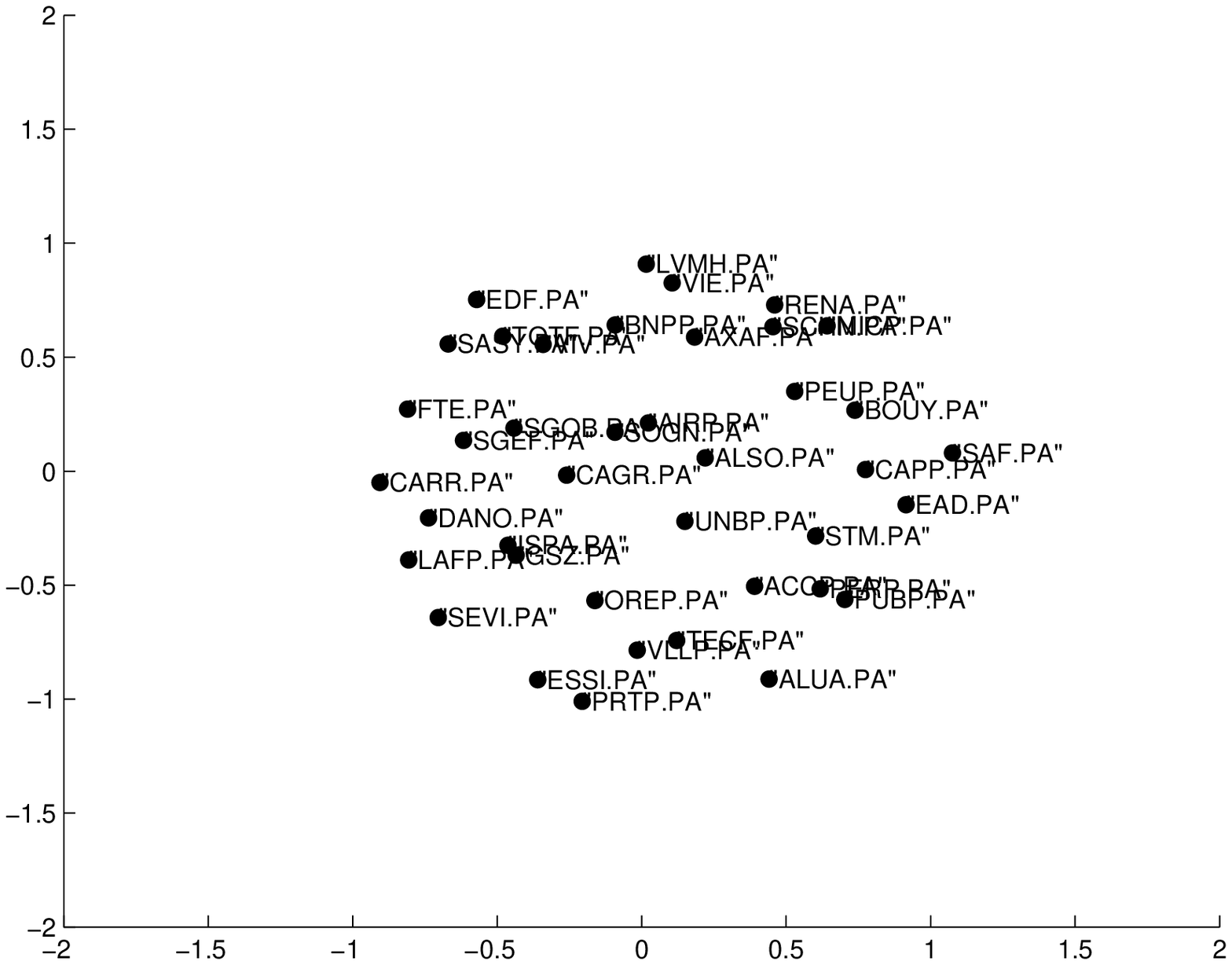}
\end{center}
\caption{MDS plots  for bins 7-12. Each point on a plot represents a stock (see list of CAC40 stocks in Table \ref{rics}), designated by two coordinates ($x_i,y_i$), $i=1, \dots, N$. We took the average of the correlation coefficients for each pair over all 22 days, and then used them to generate the MDS plot for a particular bin. We then plotted the MDS maps for the different bins to see the evolution during the day.}
\label{MDS_avg_correlations2}
\end{figure}

We further plotted the variation of the mean distance of all the coordinates from the centre of the map, over the different bins to see the temporal evolution during the day, in \fg{mean_distance1}. This follows exactly the opposite trend of the average correlations as shown in \fg{eigen} or \fg{average}-- the mean distance \textit{decreases} during the day. This result is as expected, and not very surprising.

\begin{figure}[H]
\begin{center}
\includegraphics[width=2.5in]{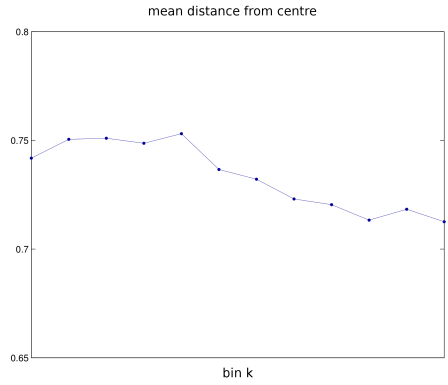}
\end{center}
\caption{Mean distance of coordinates of all the points (40 stocks) from center of the map, as a function of the bin $k$. There are 12 bins of 30 minutes between 10:00 and 16:00 CET.}
\label{mean_distance1}
\end{figure}

\begin{table}[H]

\caption{RICS list of the stocks in the CAC 40.}
\label{rics}
\begin{center}
\begin{tiny}
\begin{tabular}{|c|c|}

\hline 
Names & RICS\tabularnewline
\hline 
\hline 
ACCOR FICTIVE  & ACCP.PA \tabularnewline
\hline 
AIR LIQUIDE  & AIRP.PA \tabularnewline
\hline 
ALCATEL LUCENT  & ALUA.PA \tabularnewline
\hline 
ALSTOM  & ALSO.PA \tabularnewline
\hline 
ARCELOR MITTAL FICTIVE & ISPA.AS \tabularnewline
\hline 
AXA  & AXAF.PA \tabularnewline
\hline 
BNP PARIBAS  & BNPP.PA \tabularnewline
\hline 
BOUYGUES  & BOUY.PA \tabularnewline
\hline 
CAP GEMINI  & CAPP.PA \tabularnewline
\hline 
PERNOD RICARD  & PERP.PA \tabularnewline
\hline 
VALLOUREC  & VLLP.PA \tabularnewline
\hline 
CARREFOUR  & CARR.PA \tabularnewline
\hline 
PEUGEOT SA  & PEUP.PA \tabularnewline
\hline 
VEOLIA ENVIRONNEMENT  & VIE.PA \tabularnewline
\hline 
CREDIT AGRICOLE SA  & CAGR.PA \tabularnewline
\hline 
PPR  & PRTP.PA \tabularnewline
\hline 
VINCI & SGEF.PA \tabularnewline
\hline 
DANONE  & DANO.PA \tabularnewline
\hline 
PUBLICIS  & PUBP.PA \tabularnewline
\hline 
VIVENDI  & VIV.PA \tabularnewline
\hline 
EADS PEA FICTIVE  & EAD.PA \tabularnewline
\hline 
RENAULT  & RENA.PA \tabularnewline
\hline 
EDF  & EDF.PA \tabularnewline
\hline 
SAINT GOBAIN  & SGOB.PA \tabularnewline
\hline 
ESSILOR INTERNATIONAL  & ESSI.PA \tabularnewline
\hline 
SANOFI  & SASY.PA \tabularnewline
\hline 
FRANCE TELECOM  & FTE.PA \tabularnewline
\hline 
SCHNEIDER ELECTRIC SA  & SCHN.PA \tabularnewline
\hline 
GDF SUEZ  & GSZ.PA \tabularnewline
\hline 
SOCIETE GENERALE  & SOGN.PA \tabularnewline
\hline 
LOREAL  & OREP.PA \tabularnewline
\hline 
STMICROELECTRONICS PEA FICTIVE & STM.PA \tabularnewline
\hline 
LVMH  & LVMH.PA \tabularnewline
\hline 
SUEZ ENVIRONNEMENT SA  & SEVI.PA \tabularnewline
\hline 
LAFARGE  & LAFP.PA \tabularnewline
\hline 
TECHNIP  & TECF.PA \tabularnewline
\hline 
MICHELIN  & MICP.PA \tabularnewline
\hline 
TOTAL  & TOTF.PA \tabularnewline
\hline 
NATIXIS  & CNAT.PA \tabularnewline
\hline 
UNIBAIL-RODAMCO SE & UNBP.PA\tabularnewline
\hline 
\end{tabular}
\end{tiny}
\end{center}
\end{table}

\subsection{MDS using daily data}

In order to capture the co-movement of stocks visually, we again used the MDS plots of 54 stocks from Yahoo daily data, for the period of January 2008-May 2011. We computed the correlations using non-overlapping windows of $T$ consecutive trading days, using \eq{eq:coeff}. The choice of $T$ is important because if $T/N$ is small, then according to the Random Matrix Theory we cannot distinguish between noise and the true signal. Since MDS needs a full rank correlation matrix, the noise needs to be cleaned with appropriate statistical measures before applying MDS.

As before, using the correlation matrices as input, we made the distance transformations (using \eq{distance}) to produce the distance matrices. These distance matrices were then used as inputs to the MDS code in \texttt{MATLAB}. We used the method of simulated annealing to optimize the cost function of a particular day. The first day (time-step) starts with an initial set of coordinates chosen at random; for the following days (time-steps), we used the final results of the previous day (time-step) as the initial state\footnote{This is to avoid too drastic a change in the MDS plots from one time step to another, keeping in mind that the vectors $x_i$ are \textit{not unique}-- with the Euclidean metric, they may be arbitrarily \textit{translated} and \textit{rotated}. We imposed a small penalty in the cost function for deviation from the initial state.}. The output of the MDS were the coordinates, which were plotted as the MDS maps. The coordinates were plotted in a manner such that the centroid of the map coincided with the origin $(0,0)$. We then computed the mean distance of all the coordinates from the centre, and plotted this measure as a  functon of time.

In \fg{MDS_daily1} we plot MDS maps for sample dates: 28/05/2008 (pre-Subprime crisis), 27/10/2008 (onset of Subprime crisis) and 28/06/2010 (post-Subprime crisis). In these plots we do see the difference in the positions of the companies. The position of Lehman brothers in the plot of the MDS during the post-Subprime crisis is noteworthy. 

We also plot in \fg{MDS_daily1}, the mean distance of coordinates from center for the period 01/01/2008 to 31/12/2009. There is certainly a noticeable variation in this entire period, and the period of the Subprime crisis can be identified with the low value of mean distance.

\begin{figure}[H]
\begin{center}
\includegraphics[width=0.55\linewidth]{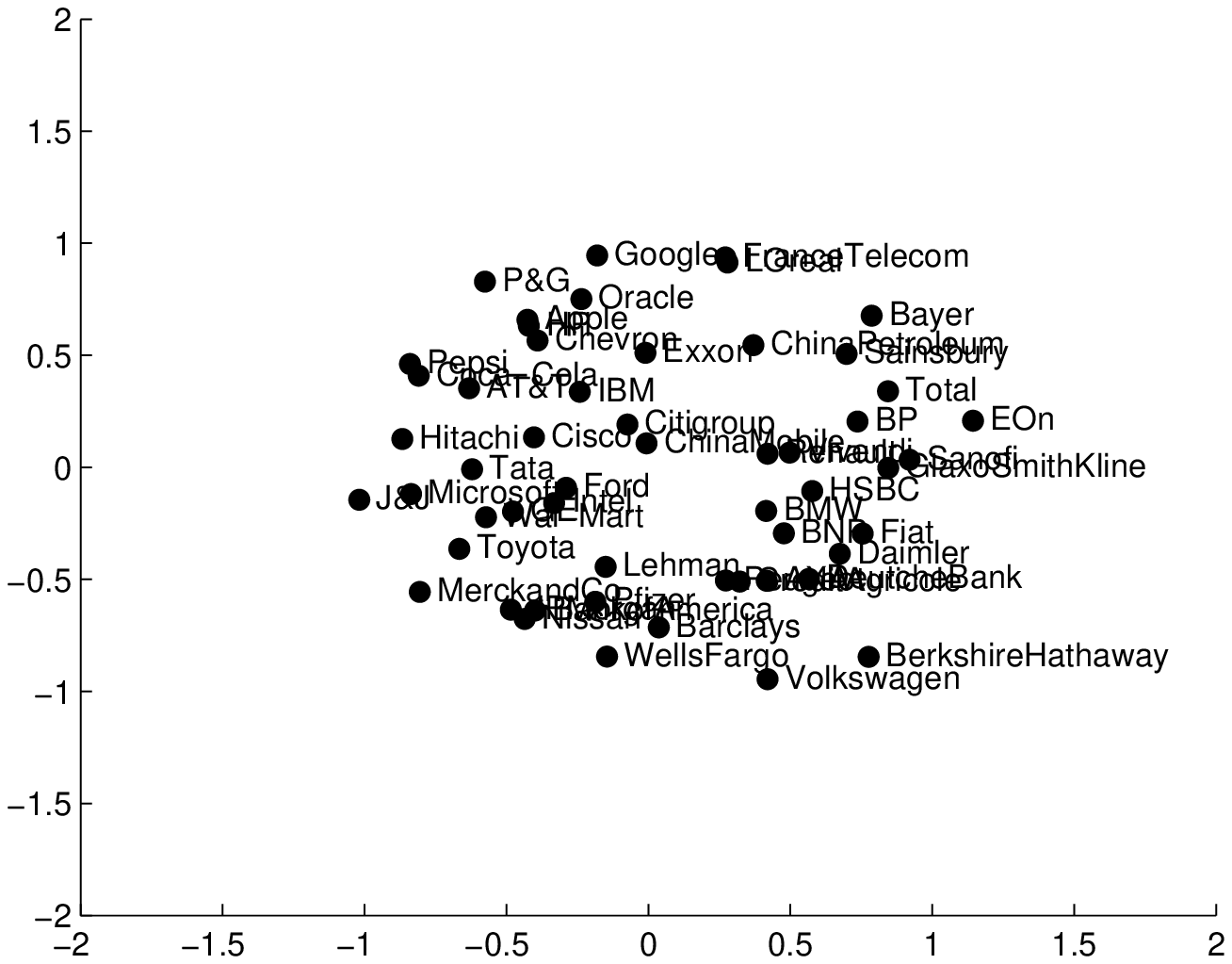}\\
\includegraphics[width=0.55\linewidth]{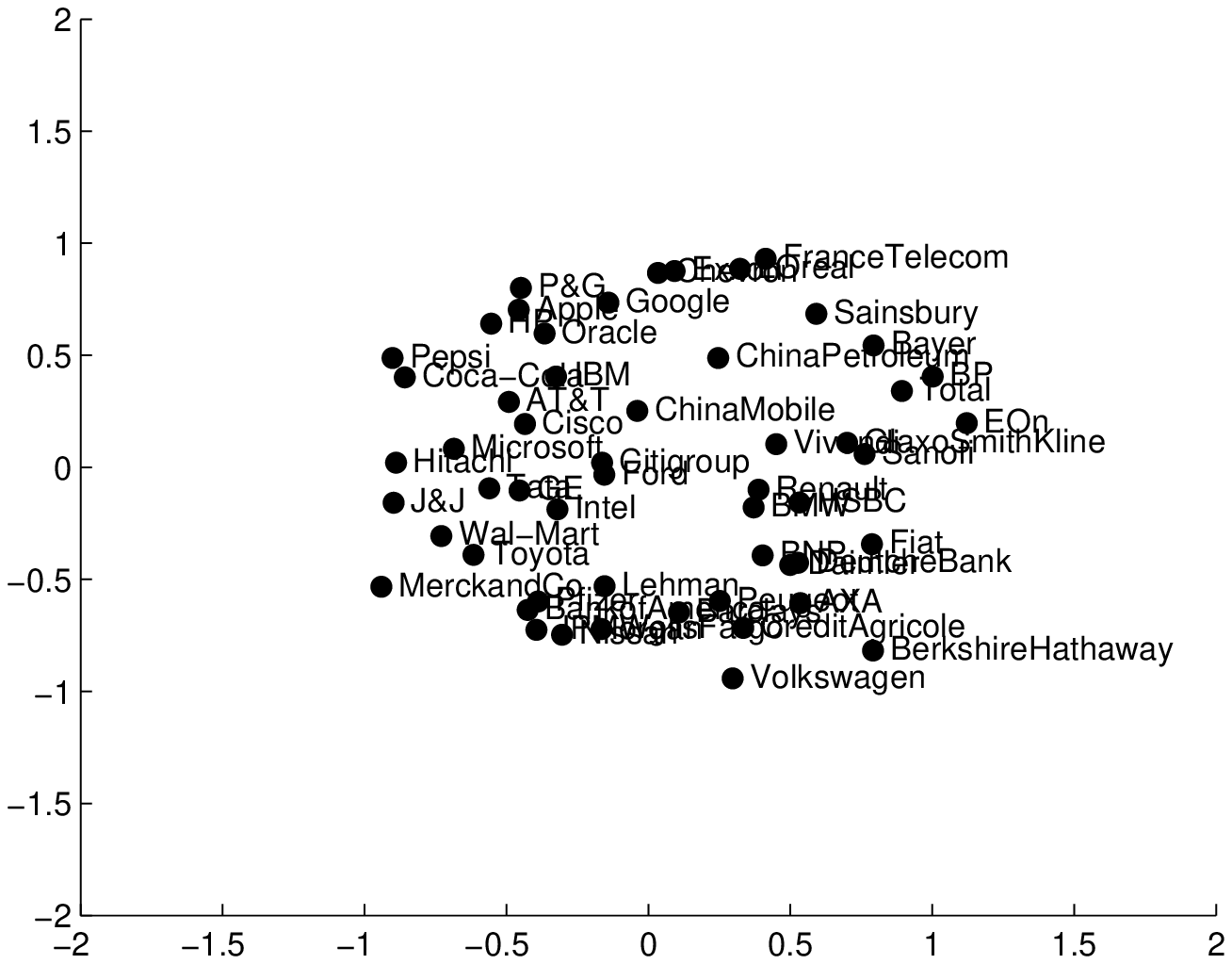}\\
\includegraphics[width=0.55\linewidth]{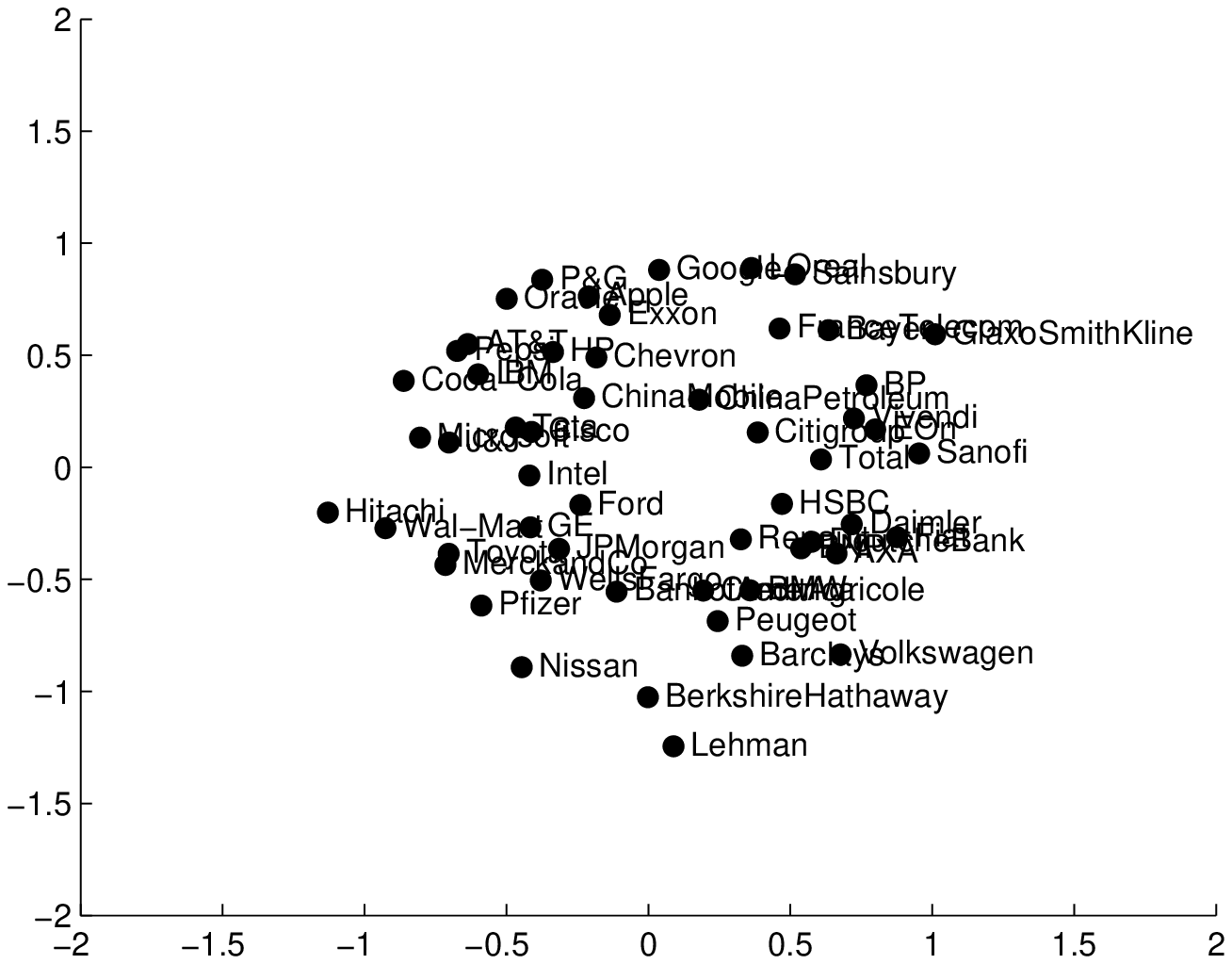}\\
\includegraphics[width=0.65\linewidth]{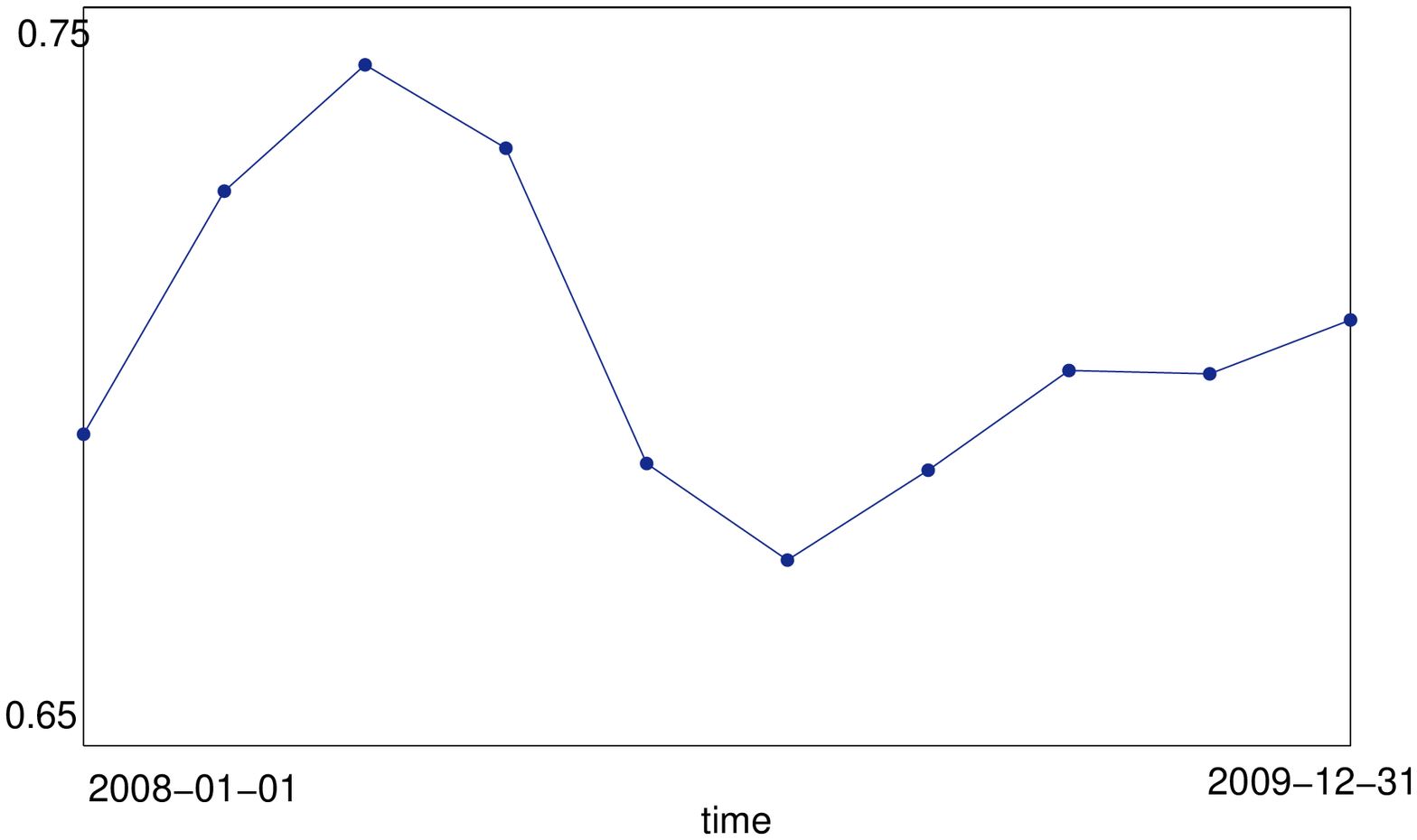}
\end{center}
\caption{The correlation matrices are computed from Yahoo daily closure price data using \eq{eq:coeff} and 54 trading day window, for the set of 54 companies. The points on each MDS plot represent stocks, each designated by two coordinates ($x_i,y_i$), $i=1, \dots, 54$. Top-most: MDS plot for date 28/05/2008. Top: MDS plot for date 27/10/2008. Bottom: MDS plot for date 28/06/2010. Bottom-most: Mean distance of coordinates from center for the two year period 01/01/2008 to 31/12/2009. }
\label{MDS_daily1}
\end{figure}

In order to examine carefully whether any clusters can be identified, we worked with a subset of 18 companies. In \fg{MDS_daily2} and \fg{MDS_daily3}, we plot MDS maps for different sample dates: 03/06/2008, 25/07/2008, and 05/09/2008 (pre-Subprime crisis); 17/10/2008, 28/11/2008 and 13/01/2009 (during Subprime crisis); 24/02/2009, 07/04/2009, 12/09/2009 and 04/11/2009 (post-Subprime crisis). 

\begin{figure}[H]
\begin{center}
\includegraphics[width=0.49\linewidth]{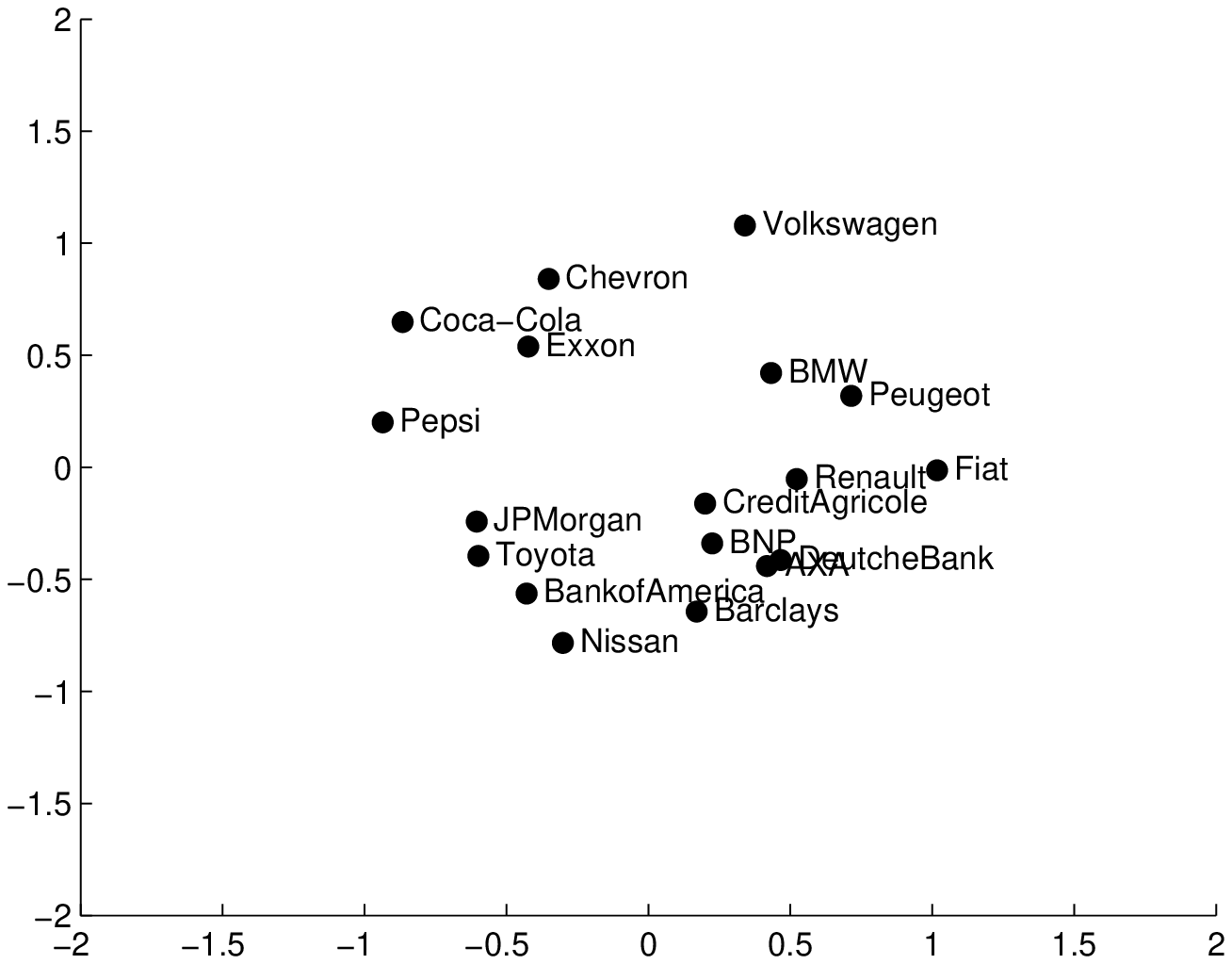}
\includegraphics[width=0.49\linewidth]{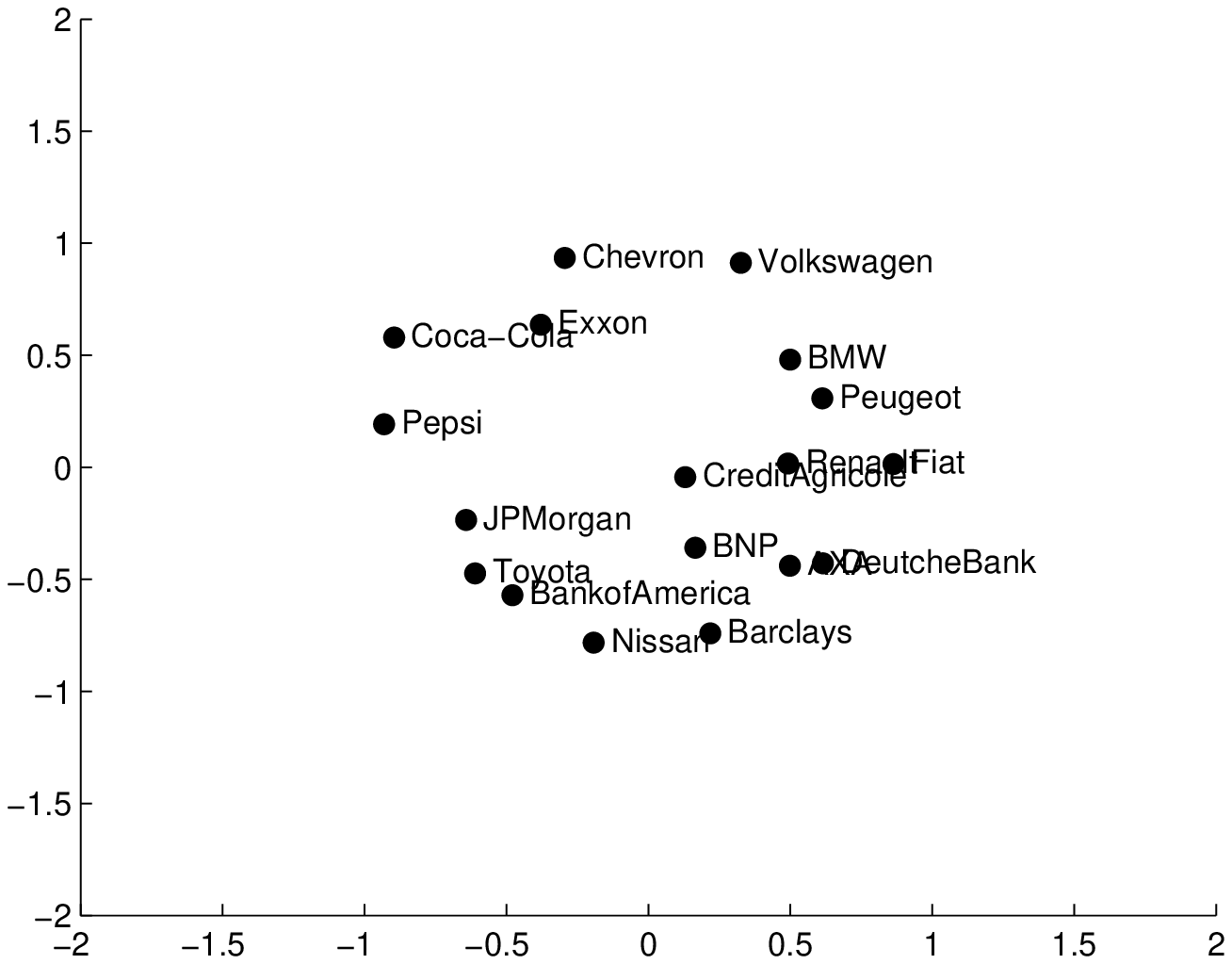}\\
\includegraphics[width=0.49\linewidth]{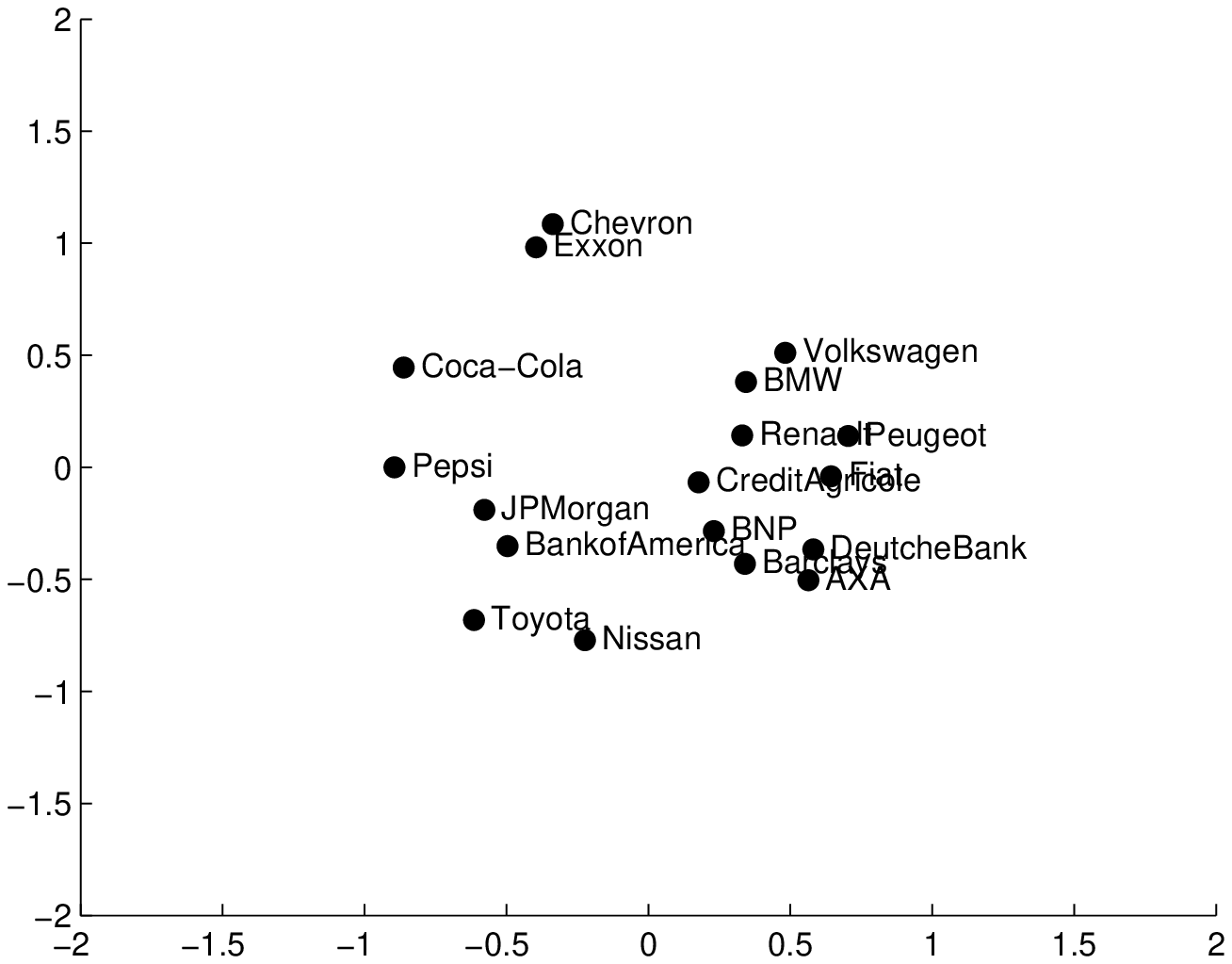}
\includegraphics[width=0.49\linewidth]{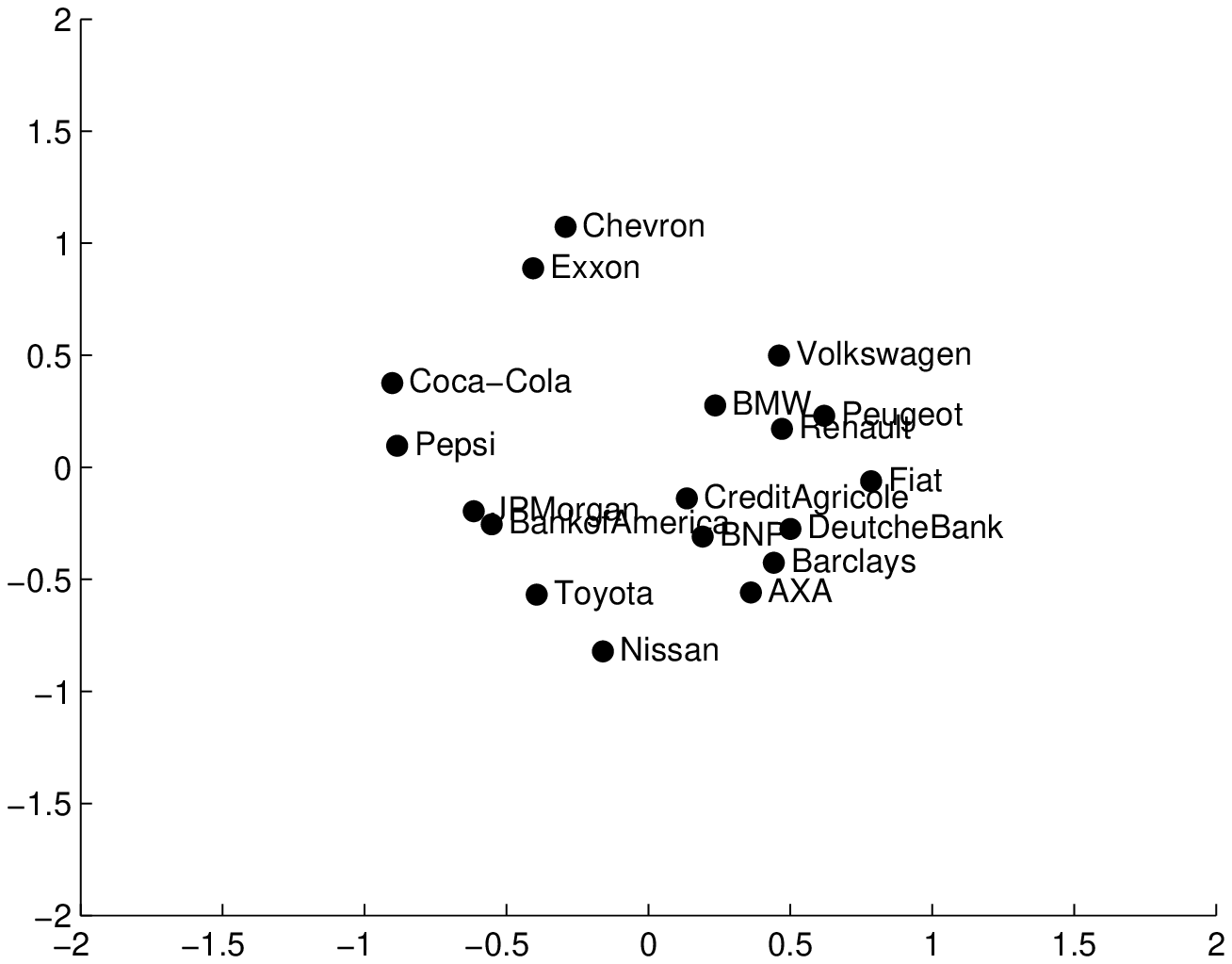}\\
\includegraphics[width=0.49\linewidth]{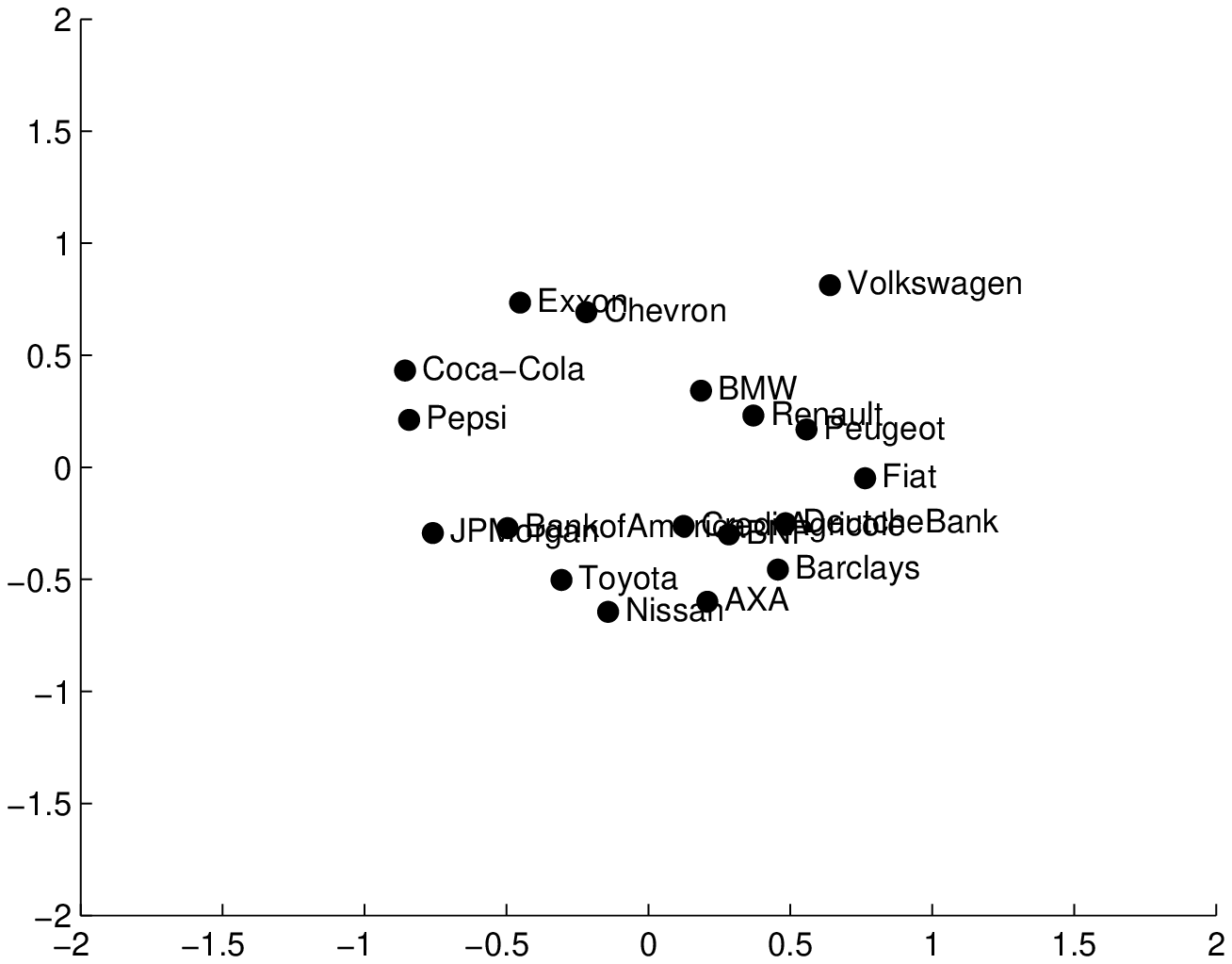}
\includegraphics[width=0.49\linewidth]{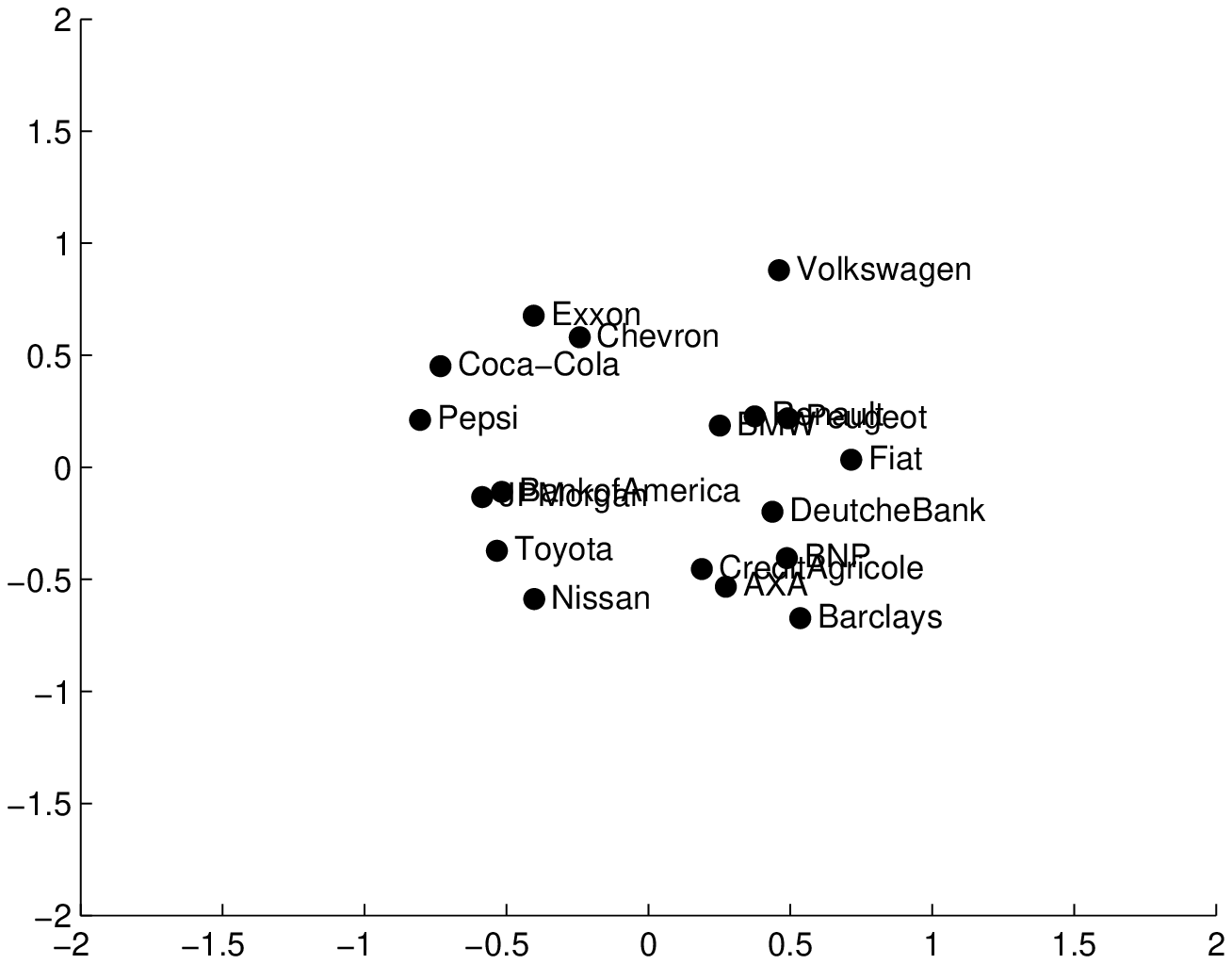}
\end{center}
\caption{MDS plots for different dates. Top Left: 03/06/2008 Top Right: 25/07/2008 
Middle Left: 05/09/2008 Middle Right: 17/10/2008
Bottom Left: 28/11/2008 Bottom Right: 13/01/2009. The correlation matrices are computed from Yahoo daily closure price data using \eq{eq:coeff} and 30 trading day window, for the subset of 18 companies. The points on each plot represent stocks, each designated by two coordinates ($x_i,y_i$), $i=1, \dots, 18$.}
\label{MDS_daily2}
\end{figure}

\begin{figure}[H]
\begin{center}
\includegraphics[width=0.49\linewidth]{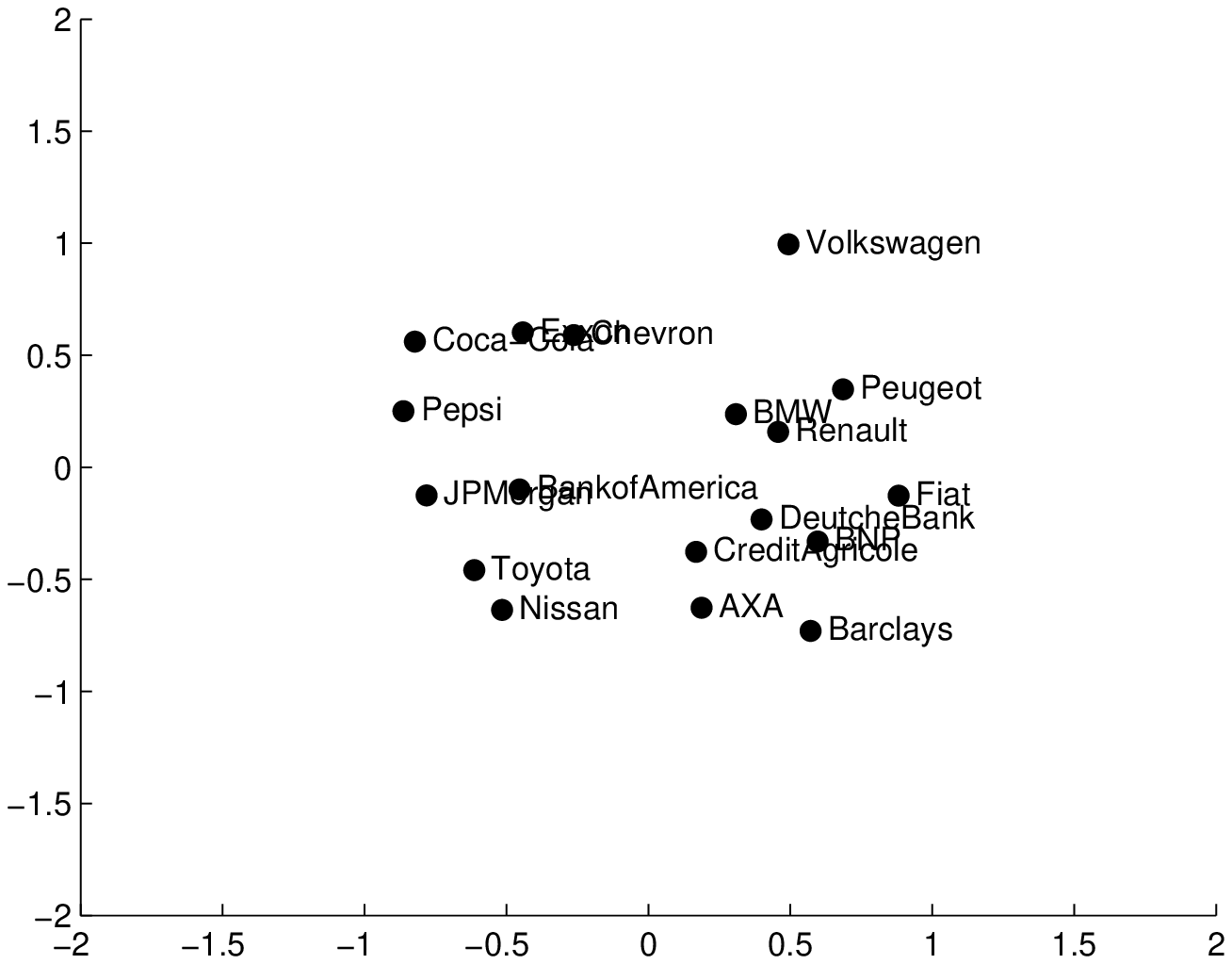}
\includegraphics[width=0.49\linewidth]{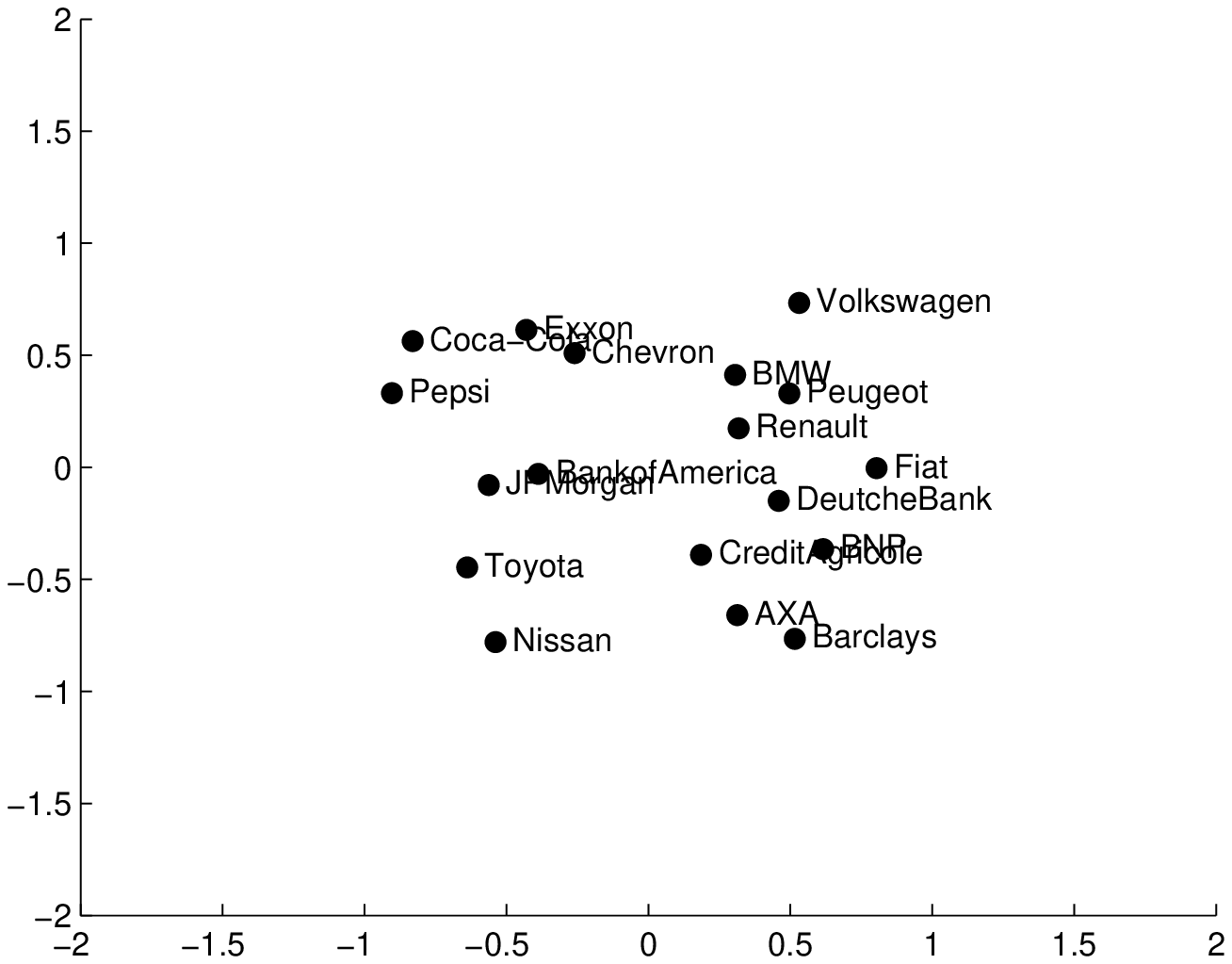}\\
\includegraphics[width=0.49\linewidth]{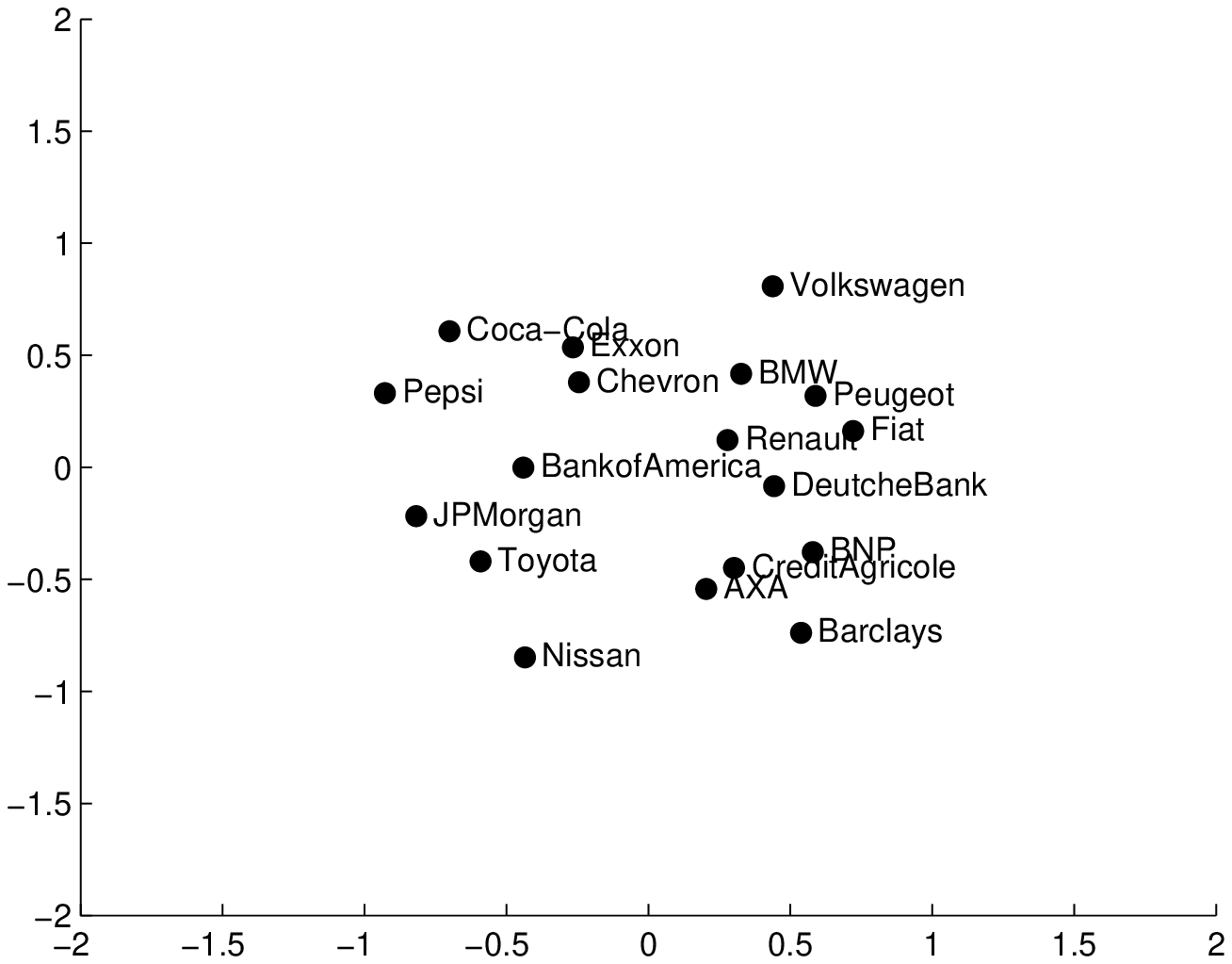}
\includegraphics[width=0.49\linewidth]{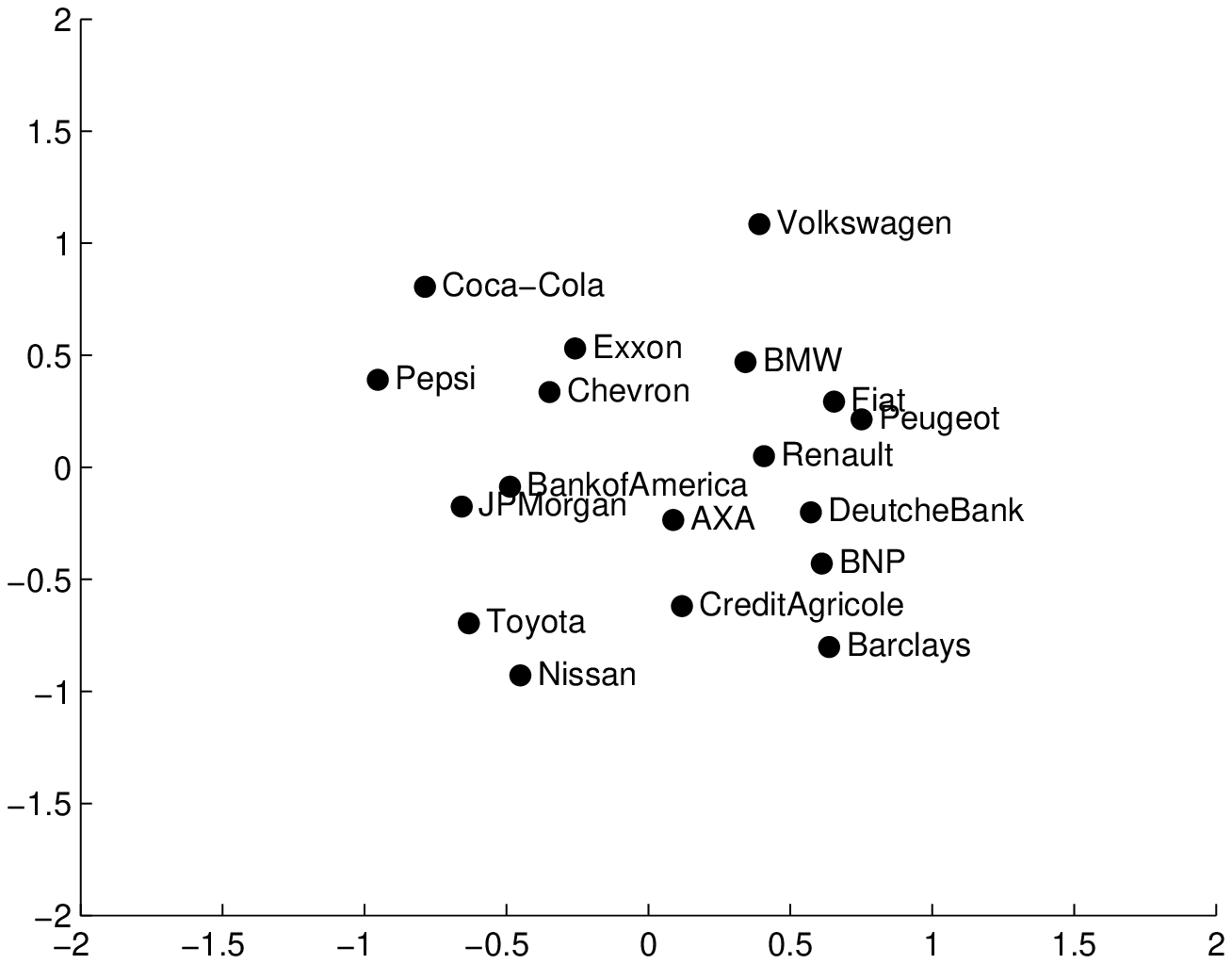}\\
\end{center}
\caption{MDS plots for different dates. Top Left: 24/02/2009 Top Right: 07/04/2009 Bottom Left: 12/09/2009 Bottom Right: 04/11/2009. The correlation matrices are computed from Yahoo daily closure price data using \eq{eq:coeff} and 30 trading day window, for the subset of 18 companies. The points on each plot represent stocks, each designated by two coordinates ($x_i,y_i$), $i=1, \dots, 18$.}
\label{MDS_daily3}
\end{figure}

In these plots we do see the considerable differences in the positions of the companies. However, it is interesting to follow the positions of certain pairs:
\begin{enumerate}
\item JP Morgan and Bank of America
\item Nissan and Toyota
\item Chevron and Exxon
\item Pepsi and Coca Cola.
\end{enumerate}
This type of visual plot may therefore be used in identifying potential pairs of stocks for ``pairs trade''. Such a strategy monitors the performances of two historically correlated stocks: when the correlation between the two securities temporarily weakens, i.e. one stock moves up while the other moves down, the pairs trade strategy would be to \textit{short} the outperforming stock and to \textit{long} the underperforming one, betting that the ``spread'' between the two would eventually converge. Further analysis is of course necessary to devise such a strategy.

We also find that there is some noticeable clustering effect, e.g. as all the European banks are in one cluster and all the European automobiles are in another cluster. 

\section{Concluding remarks}

In this paper, we first reviewed existing results on intraday patterns concerning both individual and collective stock dynamics. We studied the cross-sectional ``dispersion'' of returns and its typical evolution during the day, and found that the average volatility is high during the market opening hours, then decreases so as to reach a minimum around lunch time, and increases again steadily until the market closes. The average of $|\mu_d(k;t)|$, which is a proxy for the ``index volatility'', also displayed a
U-shaped pattern similar to that of $\sigma(k)$.
Studying the intraday pattern of the  leading modes (eigenvalues)  evaluated using the cross-correlation matrix between stock returns, we found that the maximum eigenvalue $\lambda_1(k)$ (corresponding to the market mode or \textit{average correlation}) clearly \textit{increases} as time elapses. However, the evolution of the next six eigenvalues $\lambda_i(k)$, $i=2, \dots, 7$ showed that the amplitudes of these {\it decrease} with time. 
Then, we made additional plots of the pair-wise 
cross-correlation matrix elements and studied their typical evolution during the day.
Finally, we  used multidimensional scaling (MDS) in generating maps and visualizing the dynamic evolution of the stock market during the day. When the MDS studies were repeated  with daily data, we found that it was easier to visualize  or detect specific sectors, strongly correlated pairs and market events. We suggest that this type of plots using daily data may be used in designing strategies of ``pairs trade'' as explained earlier, or identifying clusters or detecting market trends. 

\begin{acknowledgement}
The authors acknowledge F. Abergel, N. Huth, A. Jedidi and F. Pomponio for critical discussions, and thank E. Guevara H. for \fg{figure1} and \fg{eigen} (Top). A.C. and T.S. are also grateful to R. Malla and P. Nguemo for some preliminary exercises and computations.
\end{acknowledgement}

\end{document}